\documentclass[journal]{IEEEtran}[12pt]
\ifCLASSINFOpdf
\usepackage[pdftex]{graphicx}
\graphicspath{{../pdf/}{../jpeg/}}
\DeclareGraphicsExtensions{.pdf,.jpeg,.png}
\else
\usepackage[dvips]{graphicx}
\graphicspath{{../eps/}}
\DeclareGraphicsExtensions{.eps}
\fi\usepackage{graphicx}

\usepackage{graphics}
\usepackage{epsfig}
\usepackage{epstopdf}
\usepackage{stfloats}
\usepackage[cmex10]{amsmath}
\usepackage{algorithmic}
\usepackage{array}
\usepackage{mdwmath}
\usepackage{mdwtab}
\usepackage{graphicx}
\usepackage{subfigure}
\usepackage{color}
\usepackage{amsfonts,amssymb}
\usepackage{hyperref} 
\usepackage{multirow}
\usepackage{diagbox}
\usepackage{caption}
\usepackage{makecell}
\usepackage{amsfonts,amsthm,array} 
\usepackage[ruled]{algorithm2e}

\usepackage{lineno} 
\usepackage{threeparttable}
\usepackage{bm} 
\SetKwRepeat{Do}{do}{while}

\begin{document}
	

\title{Beamforming for Secure RSMA-Aided ISAC Systems	\thanks{Manuscript received.}}

\author{Qian~Dan, 
	Hongjiang~Lei, 
	Ki-Hong~Park, 
	and 
	Gaofeng~Pan 
}

\maketitle
	
\begin{abstract}

	This work investigates the physical layer security of rate-splitting multiple access (RSMA)-aided integrated communication and sensing (ISAC) systems.
	The ISAC base station (BS) transmits signals to communicate with users in an eavesdropped scenario and to estimate the parameters of the sensed targets. 
	The research considers different sensing signals under RSMA technology and the Cram{\'{e}}r-Rao bound of the parameter estimation is utilized as the sensing metric. 
	With the channel state information (CSI) of eavesdroppers known, the transmitting beam of the BS is optimized to maximize the energy efficiency in terms of the minimum user rate and secrecy capacity considering the fairness among users and ensuring the sensing performance and communication security. 
	With the CSI of eavesdroppers unknown, the transmitting beam of the BS is designed to minimize the energy consumption for sensing and communication, and the residual power is utilized for artificial noise which is isotropically emitted to achieve interference with potential eavesdroppers.  
	To solve the non-convex problems, three iterative algorithms based on successive convex approximation and penalty function are proposed.
	The simulation results illustrate the effectiveness of proposed schemes.

\end{abstract}


\section{Introduction}
\label{sec:introduction}

\subsection{Background and Related Works}
\label{sec:Background}  		

Integrated sensing and communication (ISAC) shares spectrum and hardware among radar and communication, reduces cost, weight, and size, and improves the efficiency of spectrum, energy, and hardware \cite{LiuF2022JSAC}-\cite{Prelcic2024PROC}. 
Sharing spectrum, hardware platforms, and transmitted waveforms between communication and sensing result in spectral, energy, and hardware efficiency (integration gain). 
At the same time, through careful design, the communication and sensing functions can mutually be assisted and supported to improve each other's performance so as to obtain coordination gain \cite{CuiY2021Net}. 
Rate-splitting multiple access (RSMA) is an outstanding technology to deal with inter-user interference by flexibly bridging the two extremes of fully decoding interference and fully treating interference \cite{MaoY2022Survey, OuX2023TCOM, HanS2024TCCN}.
In particular, each user's individual message is split into a common part and a private part. 
All the common parts are combined into a common message that all the users can decode, and the private parts are independently encoded and decoded only by the intended user. 
In other words, the inter-user interference is partially decoded and partially treated as noise.
In RSMA-aided ISAC systems, RSMA technology can flexibly and robustly manage not only the inter-user interference but also the interference between communication and sensing \cite{ClerckxB2024PROC, LiuY2024PROC}. 

Based on the application scenarios, the research in RSMA-aided ISAC systems has been classified as communication-centered (CC) \cite{GuJ2024TVT, YaoB2024ARX}, sensing-centered (SC) \cite{YinL2022WCNC, ChenZ2024WCL, LiuZ2024IOT, LiuY2025TCCN}, and communication-sensing trade-off (CST) \cite{XuC2021JSTSP}-\cite{GaoP2023JSAC}.
The authors in \cite{GuJ2024TVT} investigated an RSMA-aided integrated communication and jamming system with a legitimate, an illegitimate transmitter, and multiple legitimate and illegitimate receivers. 
The sum throughput of the considered system was maximized by jointly designing the beamformers (BFs), sensing time, and common rate allocation (CRA) while detecting the existence of transmissions from the illegitimate transmitter and taking the probability of false alarm (PFA) and the detection probability (DP) constraints into account. 
In \cite{YaoB2024ARX}, the authors investigated the unmanned aerial vehicle (UAV)-enabled RSMA system wherein multiple aerial ISAC base stations (BSs) transmitted signals to multiple terrestrial communication users (CUs) and detected a terrestrial survivor coordinated with a dedicated sensing receiver. 
The user scheduling, the BFs and trajectories at BSs, and the CRA were jointly optimized to maximize the weighted sum rate (WSR) of the considered system subject to the sensing SNR constraint.
The sensing-centered design focuses on improving sensing performance under communication rate constraints.
The authors in \cite{YinL2022WCNC} investigated the ISAC satellite system with multiple downlink terrestrial CUs and a single moving target. 
The trace of the Cram{\'{e}}r-Rao Bound (CRB) matrix of the target's azimuth and elevation angles was minimized by optimizing the RSMA-based BF and the CRA considering the CUs' quality of service (QoS) and transmit power budget constraints.
In \cite{ChenZ2024WCL}, the authors investigated a reconfigurable intelligent surface (RIS)-based RSMA-aided ISAC system with multiple CUs and a target. 
The CRA, the active and passive BFs at both the BS and the RIS were jointly optimized to maximize the sensing SNR of target detection.
The authors in \cite{LiuZ2024IOT} proposed a transmissive RIS (TRIS) ISAC architecture consisting of a horn antenna, TRIS, and controller and a new sensing performance metric (named sensing QoS), which was defined as the metric summation of target detection, target localization, and target tracking. 
The CRA, the sensing, and the communication stream precoding matrix were jointly optimized to maximize the sensing QoS. 
{The authors in \cite{LiuY2025TCCN} utilized simultaneously transmitting and reflecting RIS (STAR-RIS) to improve the RSMA-aided ISAC systems. The sensing signal-to-interference-plus-noise ratio (SINR) was maximized by jointly designing the transmit BFs at BS, the reflection/refraction at the STAR-RIS, and CRA. }

\textit{The simultaneous optimization of communication and sensing performance in RSMA-assisted ISAC systems has emerged as a prominent research focus in academia.}
Considering the perfect and imperfect channel state information (CSI), the authors in \cite{XuC2021JSTSP} and \cite{LoliRC2022ARX} maximized the weighted difference between the WSR and beampattern approximation mean square error (MSE) by jointly optimizing the CRA and the precoding matrices subject to the per-antenna power constraint.
In \cite{YinY2022CL}, the authors maximized the weighted summation between the minimum achievable rate (MAR) among CUs and the maximum root of CRB by jointly designing the ISAC waveform and the CRA while taking the per-antenna power constraint into account. 
In \cite{ChenKX2024TVT}, the authors maximized the weighted sum between the MAR among CUs and the smallest eigenvalue of the Fisher information matrix (FIM) by jointly designing the ISAC waveform and the CRA while considering the per-antenna power constraint. 
In \cite{GaoP2023JSAC}, the authors investigated the Pareto optimization framework of the RSMA-aided cooperative ISAC system, and the squared position error bound (SPEB) and sum rate among CUs were utilized as the localization and communication metrics, respectively.
Both the scenarios with the weighted summation and the constrained method between the sensing and communication performance were considered.

In addition, optimizing the energy consumption of BSs is also an essential focus of RSMA-assisted ISAC systems. 
For example, the authors in \cite{GalappaththigeD2024ARX} considered an integrated sensing and backscatter communication system consisting of an ISAC BS, multiple CUs, tags, and readers. 
The overall transmission power at the BS was minimized by jointly optimizing the BFs, the tag reflection coefficients, and CRA, considering the communication and sensing constraints.
Moreover, in \cite{DizdarO2022OJCS}, considering low-resolution digital-to-analog converters (DACs), the energy efficiency (EE) of the RSMA-aided ISAC systems was minimized by optimizing the precoding matrices, the number of active RF chains, and the CRA for the scenarios with perfect and imperfect CSI. 
Moreover, the authors in \cite{LiuZ2024TWC} investigated the RSMA-assisted low earth orbit satellite system with low-resolution DACs, and the minimum EE among CUs was maximized by optimizing the precoders subject to the power consumption of each RF chain constraint, the communication and sensing constraints.

\textit{In RSMA systems, the common and private streams are coded with superposition coding scheme and different power levels, which 
can mislead the external eavesdroppers through the common stream can be utilized as artificial noise (AN) \cite{SalehkalaibarS2013TIFS, FuH2020TWC}.}
For example, considered imperfect CSI, the authors in \cite{FuH2020TWC} maximized the minimum total secrecy rate (TSR) (the sum of secrecy rate of common and private streams) among the legitimate users (LUs) by carefully designing the BFs for common and private streams. 
The authors in \cite{XiaH2023CL} studied the secrecy performance of the RSMA systems with multiple legitimate and illegitimate receivers. 
The minimum total rate among all the LUs was maximized by jointly optimizing transmit BF, AN vector, and CRA subject to the SNR constraint.
The authors in \cite{GaoY2023CL} utilized RIS technology to enhance the secrecy performance of the RSMA system. 
The minimum TSR among all the LUs was maximized by jointly designing the transmit beamforming vectors (including common, private stream, and AN), the RIS's BF, and the secrecy CRA.
In \cite{XiaH2024TWC}, the authors investigated the secrecy performance of the RSMA systems with internal potential eavesdroppers. 
The weighted sum common rate and secrecy rate (WSCSR) of all the LUs with perfect CSI and the ergodic WSCSR with imperfect CSI of all users were maximized subject to the secrecy rate constraint, respectively.
In \cite{ChangH2024TGCN}, the authors utilized STAR-RIS to enhance the covert performance of RSMA systems. The transmit BF at the BS, reflection/refraction BFs at the STAR-RIS, and CRA were jointly optimized to maximize the covert rate. 

\begin{table*}
	\centering
	\caption{Comparison of Related Works on ISAC}
	
	\begin{threeparttable}
		\resizebox{0.8\textwidth}{!}
		{
			\begin{tabular}{c|c|c|c|c|c|c|c}
				\Xhline{1.2pt}
				\textbf{Ref.}&  \textbf{\makecell[c]{Communication \\metric}} & \textbf{\makecell[c]{Sensing \\metric}} &  \textbf{\makecell[c]{Optimization\\ objective}}& \textbf{\makecell[c]{Optimization \\parameters}} & \textbf{\makecell[c]{System}}  & PLS & Fairness \\
				\hline
				
				\cite{GuJ2024TVT} &  \makecell[c]{ Achievable rate }& \makecell[c]{ FAP and DP }        &  Sum throughput   &  \makecell[c]{BFs, sensing\\ time, and CRA} & \makecell[c]{CC} & & \\					
				
				\hline
				\cite{YaoB2024ARX}  &   SINR   & Sensing SNR &WSR   &  \makecell[c]{User scheduling, \\mathbfs and trajectories \\at BSs, and CRA}    & \makecell[c]{CC} & &\\		
				
				\hline
				\cite{YinL2022WCNC}   &   Achievable rate   & CRB     &   CRB 	&   \makecell[c]{ BFs and CRA}  &\makecell[c]{SC} & &\\
				
				\hline
				\cite{ChenZ2024WCL} &   Achievable rate   & Sensing SNR     &   Sensing SNR		&   \makecell[c]{ BFs at BS \\and RIS and CRA}  &\makecell[c]{SC} & & \\
				
				\hline
				\cite{LiuZ2024IOT} &   Achievable rate   & \makecell[c]{Sensing QoS  }     & \makecell[c]{ Sensing QoS }   &   \makecell[c]{Precoding matrices \\and CRA}  &\makecell[c]{SC } &&\\
				
				\hline
				\cite{LiuY2025TCCN}& Achievable rate  & \makecell[c]{Sensing SINR }     & \makecell[c]{Sensing SINR}  &     \makecell[c]{Transmit BF at the BS, \\reflection/refraction BFs at \\the STAR-RIS, and CRA}&  SC &&\\
				
				\hline
				\cite{XuC2021JSTSP, LoliRC2022ARX} &  WSR  & \makecell[c]{MSE of \\beampattern   }      &  \makecell[c]{Weighted difference \\between WSR \\and MSE  } &   \makecell[c]{Precoding matrices \\and CRA}  & \makecell[c]{CST } &&\\	
				
				\hline
				\cite{YinY2022CL}  & Achievable rate    & CRB       &  \makecell[c]{ Weighted sum \\between MAR \\and root of CRB }    & \makecell[c]{Precoding matrices \\and CRA}   &  \makecell[c]{CST } & & \checkmark \\
				
				\hline
				\cite{ChenKX2024TVT} &  \makecell[c]{Achievable rate }  & CRB    &\makecell[c]{Weighted sum between \\MAR and smallest \\eigenvalue of FIM }   &    \makecell[c]{Precoding matrices \\and CRA}  &\makecell[c]{CST } & & \checkmark\\				
				
				\hline
				\cite{GaoP2023JSAC}  &  Achievable rate   & SPEB       &  \makecell[c]{Sum-rate; SEPB; and \\weighted sum of \\SPEB and sum-rate}   &  \makecell[c]{ BFs, CRA, \\ and BS scheduling }& \makecell[c]{ CC; SC; \\and CST} && \\	
				
				\hline
				\cite{GalappaththigeD2024ARX}    & Achievable rate & \makecell[c]{Sensing rate }      &  \makecell[c]{BS transmit power } &   \makecell[c]{BFs, tag reflection \\coefficients, and CRA}  & \makecell[c]{CST } &&\\	
				
				\hline
				\cite{DizdarO2022OJCS}&  Sum rate   & \makecell[c]{MSE of \\beampattern }      & \makecell[c]{ EE }  &     \makecell[c]{ Precoding matrix, \\quantization, and \\selection matrix of \\RF chains }  &\makecell[c]{ CC } &&\\
				
				\hline
				\cite{LiuZ2024TWC} &  \makecell[c] {Achievable rate}  & CRB       &  \makecell[c]{ EE  }    & \makecell[c]{Precoding matrices and CRA}   &  \makecell[c]{CC} & & \checkmark \\
				
				\hline
				\cite{FuH2020TWC}  &   TSR  &  -     & TSR   &  \makecell[c]{ Precoding matrices }& \makecell[c]{ - } &\checkmark & \checkmark\\	
				
				\hline
				\cite{XiaH2023CL}  &   Achievable rate  &  -     & Achievable  rate   &  \makecell[c]{ Precoding matrices,\\ AN vector, and CRA }& \makecell[c]{ - } &\checkmark & \checkmark\\	
				
				\hline
				\cite{GaoY2023CL} &  \makecell[c]{TSR }  &   -      &\makecell[c]{TSR  }   &    \makecell[c]{ Transmit BFs, RIS's \\BF, and secrecy CRA }  &\makecell[c]{ - } &\checkmark & \checkmark\\								
				
				\hline
				\cite{XiaH2024TWC} &  WSR   &  -        &  \makecell[c]{ WSCSR and \\ergodic WSCSR }    &\makecell[c]{Precoding matrices \\and CRA }   &  \makecell[c]{ -} &\checkmark &\\
				
				\hline
				\cite{ChangH2024TGCN}  &  \makecell[c]{Covert rate}  & -   &\makecell[c]{   Covert rate  }   &    \makecell[c]{Transmit BFs,\\ reflection/refraction \\mathbfs, and CRA }  &\makecell[c]{ - } &\checkmark &\\				
				
				\hline
				\cite{ChuJJ2023TVT}  &  SINRs  & Sensing SINR      & \makecell[c]  {eavesdropping  SINR/\\communication and \\sensing power}   &  \makecell[c]{ Communication and \\sensing BFs}& \makecell[c]{ CC} & \checkmark&\\	
				
				\hline
				\cite{WeiW2024TWC}  &  SSR  & \makecell[c]{Beampattern }      &  \makecell[c]{SSR } &   \makecell[c]{Transmit BFs at the BS and \\STAR-RIS’s BFs }  & \makecell[c]{CC} &\checkmark &\\	
				
				\hline
				\cite{SuN2024TWC}  & Secrecy rate   & \makecell[c]{CRB}      & \makecell[c]{ Weighted sum \\normalized CRB \\and secrecy rate }  &     \makecell[c]{ BFs }  &\makecell[c]{CST } &\checkmark & \\

				\hline
				\cite{ZhaoB2025WCL}   &Secrecy rate  &\makecell[c] {Beampattern }      &\makecell[c]    {Weighted SSCR \\and Beampattern}  &  \makecell[c]{ Precoding matrices \\and secrecy CRA }& \makecell[c]{CST } &\checkmark & \\

				\hline
				\cite{ZhangC2024WCL} &  \makecell[c]{Secrecy rate}   & beampattern      &\makecell[c]{  Secrecy rate  }   &    \makecell[c]{Precoding matrices \\and secrecy CRA }  &\makecell[c]{ CC} &\checkmark & \checkmark \\	
				
				\hline
				\cite{LiuZ2024ARX}  & \makecell[c]{Secrecy rate \\and OP } & \makecell[c]{DP and MSE \\of beampattern   }     &  \makecell[c]{ TSR}    & \makecell[c]{Precoding matrices, \\time allocation, \\and secrecy CRA}  &  \makecell[c]{CC} &\checkmark & \\
				
				\hline		
				Our work  & URPR, USRPR &  CRB  & \makecell[c]{URPR; USRPR; NPC} &\makecell[c]{BFs and CRA; BFs \\and secrecy CRA.}  & \makecell[c]{ CC}   &\checkmark & \checkmark \\
				\Xhline{1.2pt}
		\end{tabular}}
	\end{threeparttable}
	\label{table1}
\end{table*}

\textit{ISAC introduces distinct security challenges due to shared spectrum utilization and the broadcast characteristics of wireless channels. Embedding information messages within radar signals exposes communications to potential eavesdropping by the sensing targets \cite{WeiZ2022Mag}.}
In \cite{ChuJJ2023TVT}, the authors investigated the secrecy performance of the communication and radar coexistence system with multiple CUs and multiple eavesdroppers. 
For the scenarios with the eavesdropping CSI, the maximum SINR among all the eavesdroppers was minimized by jointly designing the 
transmit BF at communication BS, the transmit BF, and the receive filter at the radar subject to the QoS CUs, the SINR requirements of the radar, and the transmit power constraints. 
For the scenarios without the eavesdropping CSI, both the BS and the radar utilized the redundant power to transmit AN to suppress the eavesdroppers, and the transmission power utilized for communication and sensing was minimized by jointly optimizing 
the transmit BF and the covariance of the AN vector at the BS, 
the transmit BF, the covariance of the AN vector, and the receive filter at the radar
ensuring the power constraints and the SINR requirements for communication and sensing. 
In \cite{WeiW2024TWC}, the authors utilized STAR-RIS to enhance the secrecy performance of the nonorthogonal
multiple access-aided ISAC systems wherein the target was a potential eavesdropper. 
Assuming the eavesdropping CSI was available, they maximized the secrecy sum-rate (SSR) of the considered systems by jointly designing the transmit BFs and the reflection/refraction BFs at the STAR-RIS while taking the requirement of the beampattern gain into account. 
Considering the scenarios without the knowledge of the eavesdropper, the authors in \cite{SuN2024TWC} proposed a method to estimate the amplitudes and angles of targets (potential eavesdroppers), then the weighted normalized secrecy rate and the determinant of the FIM was maximized by jointly designing the covariance matrix and BF while taking the power budget into account. 
Although all the results in these outstanding works can not be directly applied to the RSMA-aided ISAC system, they lay a solid foundation for investigating the secrecy performance of the RSMA-aided ISAC system.

{It is worth noting that both the common stream in RSMA systems and the sensing signals in ISAC systems can be utilized as AN to suppress the eavesdroppers. 
In \cite{ZhaoB2025WCL}, the precoding matrix and the CRA were jointly optimized to maximize the weighted summation secrecy common rate (SSCR) and the beampattern while taking the secrecy private rate constraint, the beampattern difference requirement in different directions, and the per-antenna power constraint into account.
In \cite{ZhangC2024WCL}, RIS was utilized to improve the secrecy performance of the RSMA-aided ISAC systems. 
Considering the imperfect CSI, the authors maximized the minimum TSR among all the LUs by jointly optimizing the communication and sensing BFs, secrecy CRA subject to the communication power budget and sensing power requirement.}
In \cite{LiuZ2024ARX}, the TRIS was utilized to enhance the RSMA-aided ISAC system with target detection and channel estimation, sensing and communication, and target tracking slots. 
Considering scenarios with imperfect CSI, the TSR of the considered systems was maximized by jointly optimizing the BF matrices of the common and private streams, the time allocation, the outage probability (OP) of reliability and intercept.

\subsection{Motivation and Contributions}
\label{sec:Motivation}

In the RSMA-aided ISAC systems, the targets are sensed by utilizing common stream \cite{XuC2021JSTSP} or the specific signal \cite{YaoB2024ARX},  \cite{XuC2021JSTSP}, \cite{GalappaththigeD2024ARX}. 
For the scenarios with external eavesdroppers, the common stream is designed as AN to enhance the security of the private streams, like \cite{XiaH2023CL, XiaH2024TWC}. 
To the best of the author’s knowledge, no open literature addresses the following questions:
\textit{ 
	For the RSMA-aided ISAC scenarios wherein security is required, how should the BFs be designed to obtain the best communication while taking fairness, sensing performance, and security into account? 
	Which will obtain the better fair secrecy performance in utilizing the common stream or the extra signal for sensing?
	}
To answer these questions, we investigated the performance of the RSMA-aided ISAC systems with multiple LUs, eavesdroppers, and targets. Different signals are utilized for sensing, and the BFs are designed for scenarios with and without eavesdropping CSI.
The contributions of this paper are summarized as follows. 
\begin{enumerate}
	\item We investigated the performance of the RSMA-aided ISAC systems with multiple LUs, eavesdroppers, and targets. Two schemes are proposed by utilizing the extra signal or the common stream for sensing. For the scenarios with eavesdropping CSI, BFs and CRA are jointly designed for two different objectives: one is to maximize the minimum user's rate to BS's power ratio (URPR), and the other is to maximize the minimum user's secrecy rate to BS's power ratio (USRPR). For the scenarios without the eavesdropping CSI, the BF matrices are designed to minimize the necessary power consumption (NPC) for communication and sensing, thereby improving the system's secrecy performance.
	
	\item Three iterative algorithms based on successive convex approximation (SCA) and penalty function are proposed to solve the formulated problems. Numerical results are given to validate the convergence and effectiveness of the proposed schemes. The results show that utilizing the common stream can obtain better communication performance than utilizing the extra signal for sensing.

	\item Relative to \cite{YaoB2024ARX},  \cite{XuC2021JSTSP}, \cite{GalappaththigeD2024ARX} in which the  performance of the RSMA-aided ISAC systems was investigated, we consider the problem of minimizing the transmit power at BS and maximizing the minimum performance among all the LUs subject to the QoS, security, and sensing constraints. 
	Relative to \cite{XiaH2023CL}, \cite{GaoY2023CL}, and \cite{XiaH2024TWC} in which the security of RSMA systems was investigated, the secrecy performance of RSMA-aided ISAC system is considered in this work.
	
\end{enumerate}

\begin{table}
	
	\caption{\textit{List of Notations.}}
	\begin{center}
		\begin{tabular}{c| c }
			\Xhline{1.2pt}
			\textbf{Notation}   	& \textbf{Description}								\\
			\hline
			${\alpha }$              & Path loss coefficient	\\
			\hline
			${\beta _0}$             	& Channel gain at reference distance 1 m\\
			\hline
			$\beta_t$					& \makecell[c]{Normalized coefficients for target reflection \\complex amplitude and path attenuation of the $t$-th target 	}				\\
			\hline
			${\rho _k}, {\rho _m}$         & Rician factor					\\
			\hline
			${\bm a}_r, {\bm a}_t$     & Steering vector of arrays  				\\
			\hline
			${\bm A}_r$, ${\bm A}_t$     & \makecell[c]{Steering matrix for targets' angles of the arrays  }				\\
			\hline
			$\bm B$						& \makecell[c]{Diagonal matrix of normalized matrix for \\target reflection complex
				amplitude \\and path attenuation of target angles}\\
			\hline
			$\bm \theta$   				&\makecell[c]{Vector of azimuth of the pre-estimated targets}\\
			\hline
			${\theta _k (\theta _m)}$		& Azimuth of the $k$-th LU ($m$-th eavesdropper)\\
			\hline
			$\theta_{T_k}$   &Azimuth of the pre-estimated target $T_k$\\
			\hline
			${d_{{u_k}}}({d_m})$			& \makecell[c]{Distance between $S$ and the $k$-th \\LU ($m$-th eavesdropper)	}				\\
			\hline
			${N_t} ({N_r})$&   Number of transmitting (receiving) antennas\\
			\hline
			$d$&  Antenna spacing\\
			\hline
			$\lambda $ & Wavelength\\
			\hline
			$\sigma _k^2 (\sigma _m^2)$	& \makecell[c]{ Noise power of the $k$-th LU ($m$-th eavesdropper)}\\
			\hline
			$c_k$ 						&     Messages belonging to the $k$ in the common streams  \\
			\hline
			${R_{\rm{U}}^{{\rm{th}}}}, R_{\sec }^{\rm th}$	&    Threshold of communication/security rate\\
			\hline
			${{\mathbf{h}}_k}, {{\mathbf{g}}_m}$	&  \makecell[c]{Channel between $S$ and the $k$-th \\LU ($m$-th eavesdropper)}  \\
			\hline
			${{\mathbf{w}}_c}, {{\mathbf{w}}_k}, {{\mathbf{w}}_v}$&   \makecell[c]{ BF vectors for common streams/ \\private streams for the $k$-th LU/ extra signals}  \\
			\hline
			$P$               & Transmit power of $S$ 							\\
			\hline
			$\vartheta $ 					&Threshold of the determinant of CRB			\\ 
			\hline
			$\tau $ 					& Convergence accuracy		\\ 
			\hline
			\Xhline{1.2pt}
		\end{tabular}
	\end{center}
	\label{table2}
\end{table}

\subsection{Organization}

{\color{black}
The remainder of this paper is summarized as follows. 
Sect. \ref{sec:SystemModel} introduces the system model of RSMA-aided ISAC systems.
In Sections \ref{sec:WithECSI} and \ref{sec:WithoutECSI}, the scenarios with or without eavesdropping CSI are considered respectively and the minimum UPRP, the minimum USPRP, and the NPC are maximized respectively by designing the BF matrices and efficient iterative algorithms are proposed to solve the formulated problems based on SCA and penalty function.
In Section \ref{sec:Simulation}, the numerical results of the proposed schemes are provided and analyzed.
Finally, conclusion is provided in Section \ref{sec:Conclusions}.} 
{Table \ref{table1} compares the typical works on ISAC and our work. 
	Table \ref{table2} lists the notations and symbols utilized in this work.}

\color{black}
\section{System Model}
\label{sec:SystemModel}

\begin{figure}[t]
	\centering		
	\includegraphics[width = 0.25 \textwidth]{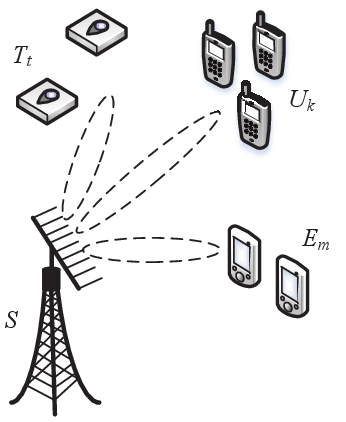}
	\caption{System model consisting of $K$ LUs, $M$ eavesdroppers,  $T$ targets, and a BS.}
	\label{fig_model}
\end{figure}

As shown in Fig. \ref{fig_model}, we consider a downlink RSMA-assisted ISAC secure communication system that is composed of an ISAC BS ($S$), $K \ge 1$ LUs $\left( {{U_k}, k = 1, \cdots ,K} \right)$, $M \ge 1$ eavesdroppers $\left( {{E_m}, m = 1, \cdots ,M} \right)$, and $T \ge 1$ point targets $\left( {{T_t}, t = 1, \cdots ,T} \right)$. 
$S$ is equipped with $ N_t$ transmit antennas to transmit communication and sensing signals and $N_r$ receive antennas to receive echo signals from $T_t$. 
All the antennas at $S$ follow a uniform linear array layout. 
All the legitimate and illegitimate receivers are equipped with a single antenna.
It is assumed that $S$ utilizes the 1-layer RSMA scheme proposed in \cite{MaoY2022Survey}  and the transmission signal is expressed as
\begin{align}
	{\mathbf{x}} = \underbrace {{{\mathbf{w}}_c}{{{s}}_c}}_{{\rm{common}}\;\,{\rm{signal}}} + \underbrace {\sum\limits_{k = 1}^K {{{\mathbf{w}}_{k}}{{{s}}_{k}}} }_{{\rm{private}}\;{\rm{signal}}} + \underbrace {{{\mathbf{w}}_{v}}{{{s}}_v}}_{{\rm{extra}}\;{\rm{signal}}}, \label{signalx}
\end{align}			
where 
${\mathbf{w}}_c$, ${\mathbf{w}}_{k}$, and ${{\mathbf{w}}_{v}}$ denote the beamforming vector of the common stream, private stream, and the extra signal respectively, 
${{{s}}_c}$ and ${{{s}}_v}$ signify the common stream and the extra signal, respectively, 
${{{s}}_k}$ is the private stream symbol for $U_k$. 
{ It is assumed that ${{{s}}_c}$, ${{{s}}_k}$, and ${{{s}}_v}$ are assumed to be zero mean and unit power.}

\subsection{Communication Model}	

{Like \cite{WeiW2024TWC}, it is assumed that the channel between $S$ and $U_k$ experiences a Rician distribution,} which is expressed as
\begin{align}
	{{\mathbf{h}}_k} = \sqrt {\frac{{{\beta _0}}}{{{{\left( {d_k} \right)}^{{\alpha }}}}}} \left( {\sqrt {\frac{{{\rho _k}}}{{1 + {\rho _k}}}} {\mathbf{\bar h}}_{\rm{LoS}}^k + \sqrt {\frac{1}{{1 + {\rho _k}}}} {\mathbf{\tilde h}}_{\rm{NLoS}}^k} \right), \label{channelhk}
\end{align}
where 
${{\mathbf{h}}_k} \in {C^{{N_t} \times 1}}$, $\beta_0$ denotes the channel gain at reference distance 1m, 
$d_k$ denotes the distance between $S$ and $U_k$, 
$\alpha $ denotes the path loss coefficient, 
{$\rho _k$ denotes the Rician factor,}
${\mathbf{\bar h}}_{{\rm{LoS}}}^k = {\mathbf{a}}({\theta _k}) = \left[ {1,{e^{j\frac{{2\pi }}{\lambda }d\sin \left( {{\theta _k}} \right)}},...,{e^{j\frac{{2\pi }}{\lambda }d\left( {{N_t} - 1} \right)\sin \left( {{\theta _k}} \right)}}} \right]$ denotes the constituents of the line-of-sight (LoS) link, 
$\theta_k$ is the azimuth angle of $U_k$, 
$\lambda$ is wavelength, 
$d $ is the adjacent antenna spacing,  
${\mathbf{\tilde h}}_{\rm{NLoS}}^k$ denotes the constituent of the non-LoS link, which obeys a complex Gaussian distribution with zero mean and unit variance.
The signal received by $U_k$ is expressed as
\begin{align}
	{y_k} &= {\mathbf{h}}_k^H{\mathbf{x}} + {n_k} \nonumber\\
	&= {\mathbf{h}}_k^H\left( {{{\mathbf{w}}_c}{{{s}}_c} + \sum\limits_{k \in K} {{{\mathbf{w}}_k}{{{s}}_k}}  + {{\mathbf{w}}_{v}}}{{{{s}}_v}} \right) + {n_k}, \label{signalyk}
\end{align}
where $n_k$ denotes additive white Gaussian noise (AWGN).

{ Like \cite{MaoY2022Survey, LiuZ2024IOT, XiaH2023CL},} 
all the LUs decode the common stream first based on a shared codebook and extracts its own information when receiving the signal stream, the private stream of all the LUs and extra signal are treated as interference. Then the SINR of the common stream at $U_k$ is expressed as
\begin{align}
	{\gamma _{c,k}} &= \frac{{{{\left| {{\mathbf{h}}_k^H{{\mathbf{w}}_c}} \right|}^2}}}{{\sum\limits_{i = 1}^K {{{\left| {{\mathbf{h}}_k^H{{\mathbf{w}}_i}} \right|}^2}}  + {{\left| {{\mathbf{h}}_k^H{{\mathbf{w}}_v}} \right|}^2} + \sigma _k^2}} \nonumber\\
	&= \frac{{{\mathbf{h}}_k^H{{\mathbf{W}}_c}{{\mathbf{h}}_k}}}{{\sum\limits_{i = 1}^K {{\mathbf{h}}_k^H{{\mathbf{W}}_i}{{\mathbf{h}}_k}}  + {\mathbf{h}}_k^H{{\mathbf{W}}_v}{{\mathbf{h}}_k} + \sigma _k^2}}, \label{SNRck}
\end{align}	
where ${\mathbf{W}}_q={\mathbf w}_q{\mathbf w}_q^H$ and $q \in \left\{ {c,k,v} \right\}$.
After decoding the common stream, $U_k$ adopts the successive interference cancellation (SIC) technique to remove the common signal from the received signal and then decode the private part. 
When decoding the private signal, all signals other than its own private signal are regarded as interference, and the SINR of the private part at $U_k$ is expressed as			
\begin{align}
	{\gamma _{p,k}} &= \frac{{{{\left| {{\mathbf{h}}_k^H{{\mathbf{w}}_k}} \right|}^2}}}{{\sum\limits_{i = 1,i \ne k}^K {{{\left| {{\mathbf{h}}_k^H{{\mathbf{w}}_i}} \right|}^2}}  +  {{\left| {{\mathbf{h}}_k^H{{\mathbf{w}}_{v}}} \right|}^2} + {\sigma_k ^2}}} \nonumber\\
	&=\frac{{ {{\mathbf{h}}_k^H{{\mathbf{W}}_k}{{\mathbf{h}}_k}} }}{{\sum\limits_{i = 1,i \ne k}^K { {{\mathbf{h}}_k^H{{\mathbf{W}}_i}{{\mathbf{h}}_k}}  +   {{\mathbf{h}}_k^H{{{{\mathbf{W}}_v}}}{{\mathbf{h}}_k}}  + {\sigma_k ^2}} }}. \label{SNRpk}
\end{align}	

{
	It must be notes that since ${{\mathbf{h}}_k}$ is random variable, ${\gamma _{c,k}}$ and ${\gamma _{p,k}}$ also are random
	variables. Thus, with the method in Refs. \cite{SuN2024TWC, HuaM2020TCOM}, and \cite{LeiH2024TVT}, the approximated SINRs are expressed as 
	\begin{align}
		{\bar \gamma _{c,k}} &= \frac{{{\mathbb{E}}\left\{ {{\mathbf{h}}_k^H{{\mathbf{W}}_c}{{\mathbf{h}}_k}} \right\}}}{{{\mathbb{E}}\left\{ {\sum\limits_{i = 1}^K {{\mathbf{h}}_k^H{{\mathbf{W}}_i}{{\mathbf{h}}_k}} } \right\} + {\mathbb{E}}\left\{ {{\mathbf{h}}_k^H{{\mathbf{W}}_v}{{\mathbf{h}}_k}} \right\} + \sigma _k^2}} \nonumber \\
		& = \frac{{{\rm{tr}}\left\{ {{{\mathbf{H}}_k}{{\mathbf{W}}_c}} \right\}}}{{{\rm{tr}}\left\{ {{{\mathbf{H}}_k}\sum\limits_{i = 1}^K {{{\mathbf{W}}_i}} } \right\} + {\rm{tr}}\left\{ {{{\mathbf{H}}_k}{{\mathbf{W}}_v}} \right\} + \sigma _k^2}} \label{barSNRck}
	\end{align}	
	and 
	\begin{align}
		{\bar \gamma _{p,k}} &= \frac{{{\mathbb{E}}\left\{ {{\mathbf{h}}_k^H{{\mathbf{W}}_k}{{\mathbf{h}}_k}} \right\}}}{{{\mathbb{E}}\left\{ {\sum\limits_{i = 1,i \ne k}^K {{\mathbf{h}}_k^H{{\mathbf{W}}_i}{{\mathbf{h}}_k}} } \right\} + {\mathbb{E}}\left\{ {{\mathbf{h}}_k^H{{\mathbf{W}}_v}{{\mathbf{h}}_k}} \right\} + \sigma _k^2}} \nonumber \\
		& = \frac{{{\rm{tr}}\left\{ {{{\mathbf{H}}_k}{{\mathbf{W}}_k}} \right\}}}{{{\rm{tr}}\left\{ {{{\mathbf{H}}_k}\sum\limits_{i = 1,i \ne k}^K {{{\mathbf{W}}_i}} } \right\} + {\rm{tr}}\left\{ {{{\mathbf{H}}_k}{{\mathbf{W}}_v}} \right\} + \sigma _k^2}}, \label{barSNRpk}
	\end{align}	
respectively, 
where 
${{\mathbf{H}}_k} = {\mathbb{E}}\left\{ {{{\mathbf{h}}_k}{\mathbf{h}}_k^H} \right\} = {a_k^0}\left( {{\mathbf{\bar h}}_{{\rm{LoS}}}^k} \right){\left( {{\mathbf{\bar h}}_{{\rm{LoS}}}^k} \right)^H} + {a_k^1}{{\mathbf{I}}_{N_t}}$, 
${a_k^0} = \frac{{{\beta _0}{\rho _k}}}{{\left( {1 + {\rho _k}} \right){{\left( {d_k} \right)}^{{\alpha }}}}}$, 
and 
${a_k^1} = \frac{{{\beta _0}}}{{\left( {1 + {\rho _k}} \right){{\left( {d_k} \right)}^{{\alpha }}}}}$.
}
Then, the common and private rates for $U_k$ are expressed as			
\begin{align}
	{R_{c,k}} = {\log _2}\left( {1 + {\bar \gamma _{c,k}}} \right) \label{Rck}
\end{align}	
and 
\begin{align}
	{R_{p,k}} = {\log _2}\left( {1 + {\bar \gamma _{p,k}}} \right),	 \label{Rpk}
\end{align}				
respectively.

{Like \cite{MaoY2022Survey, YinY2022CL, ChenKX2024TVT}, 
we define $c_k$ as the allocated common rate to $U_k$. 
Then, it has ${R_c}= \sum\limits_{k = 1}^K {{c_k}}$, where ${R_c} = \mathop {\min }\limits_{k \in {\rm K}} {R_{c,k}}$. 
The total achievable rate of $U_k$ is denoted as 
\begin{align}
	R_k^{\rm{total}} = c_k + R_{p,k}. \label{Rktotal}
\end{align}	}

The channel between $S$ and $E_m$ is expressed as
\begin{align}
	{{\mathbf{g}}_m} = \sqrt {\frac{{{\beta _0}}}{{{{\left( {{d_m}} \right)}^{{\alpha }}}}}} \left( {\sqrt {\frac{{{\rho _m}}}{{1 + {\rho _m}}}} {{\mathbf{\bar g}}}_{\rm{LoS}}^m + \sqrt {\frac{1}{{1 + {\rho _m}}}} {{\mathbf{\tilde g}}}_{\rm{NLoS}}^m} \right),
\end{align}		
where 
$d_{m}$ denotes the distance between $S$ and $E_m$,  
$\rho _m$ denotes the Rician factor, 
${\mathbf{\bar g}}_{{\rm{LoS}}}^m = {\mathbf{a}}({\theta _m}) = \left[ {1,{e^{j\frac{{2\pi }}{\lambda }d\sin \left( {{\theta _m}} \right)}},...,{e^{j\frac{{2\pi }}{\lambda }d\left( {{N_t} - 1} \right)\sin \left( {{\theta _m}} \right)}}} \right] $ signifies transmit steering vector, 
$\theta_m$ is the azimuth angle of $E_m$, 
${\mathbf{\tilde g}}_{\rm{NLoS}}^m$ denotes the constituent of the non-LoS link, which obeys a complex Gaussian distribution with a mean of zero and unit variance.
The received signal at $E_m$ is expressed as
\begin{align}	
	{y_m} = {\mathbf{g}}_m^H{\mathbf{x}} + {n_m},
\end{align}	
where $n_m$ denotes AWGN, and the SINR of the common signal received at $E_m$ is approximated as \footnote{ In this work, it is assumed that all the eavesdroppers wiretap the confident messages independently, which is widely adopted in existing literature, such as \cite{XiaH2023CL, XiaH2024TWC, ChuJJ2023TVT, SuN2024TWC}. The scenarios with multiple colluding eavesdroppers (such as \cite{LeiH2024TVT}) will be considered in future work. } 
\begin{align}
	{\bar \gamma _{c,m}^{\rm{E}}} &= \frac{{{\mathbb{E}}\left\{ {{{\left| {{\mathbf{g}}_m^H{{\mathbf{w}}_c}} \right|}^2}} \right\}}}{{{\mathbb{E}}\left\{ {\sum\limits_{i = 1}^K {{{\left| {{\mathbf{g}}_m^H{{\mathbf{w}}_i}} \right|}^2}} } \right\} + {\mathbb{E}}\left\{ {{{\left| {{\mathbf{g}}_m^H{{\mathbf{w}}_v}} \right|}^2}} \right\} + \sigma _m^2}} \nonumber\\
	&= \frac{{{\mathbb{E}}\left\{ {{\mathbf{g}}_m^H{{\mathbf{W}}_c}{{\mathbf{g}}_m}} \right\}}}{{{\mathbb{E}}\left\{ {\sum\limits_{i = 1}^K {{\mathbf{g}}_m^H{{\mathbf{W}}_i}{{\mathbf{g}}_m}} } \right\} + {\mathbb{E}}\left\{ {{\mathbf{g}}_m^H{{\mathbf{W}}_v}{{\mathbf{g}}_m}} \right\} + \sigma _m^2}} \nonumber \\
	& = \frac{{{\rm{tr}}\left\{ {{{\mathbf{G}}_m}{{\mathbf{W}}_c}} \right\}}}{{{\rm{tr}}\left\{ {{{\mathbf{G}}_m}\sum\limits_{i = 1}^K {{{\mathbf{W}}_i}} } \right\} + {\rm{tr}}\left\{ {{{\mathbf{G}}_m}{{\mathbf{W}}_v}} \right\} + \sigma _m^2}},
\end{align}	
where 
${{\mathbf{G}}_m} = {\mathbb{E}}\left\{ {{{\mathbf{g}}_m}{\mathbf{g}}_m^H} \right\} = {b_k^0}\left( {{\mathbf{\bar g}}_{{\rm{LoS}}}^m} \right){\left( {{\mathbf{\bar g}}_{{\rm{LoS}}}^m} \right)^H} + {b_k^1}{{\mathbf{I}}_{N_t}}$, 
${b_k^0} = \frac{{{\beta _0}{\rho _k}}}{{\left( {1 + {\rho _k}} \right){{\left( {d_k} \right)}^{{\alpha }}}}}$, 
and 
${b_k^1} = \frac{{{\beta _0}}}{{\left( {1 + {\rho _k}} \right){{\left( {d_k} \right)}^{{\alpha }}}}}$.
The eavesdropping rate of $E_m$ on the common stream is expressed as
\begin{align}	
	{R_{c,m}^{\rm{E}}} =  {\log _2}\left( {1 + {\bar \gamma _{c,m}}^{\rm{E}}} \right).
\end{align}		
The eavesdropping rate at $E_m$ is expressed as
\begin{align}	
	{R_{p,k,m}^{\rm{E}}} &=  {\log _2}\left( {1 + {\bar \gamma _{p,k,m}^{\rm{E}}}} \right),
\end{align}		
where 
${\bar \gamma _{p,k,m}^{\rm{E}}}$ denotes the approximated SINR at $E_m$ that wiretap $s_{k}$, 
which is expressed as (\ref{SNREpkm}), shown at the top of this page.

\begin{figure*}[ht]
	\begin{align}
		{\bar \gamma _{p,k,m}^{\rm{E}}} &= \frac{{{\mathbb{E}}\left\{ {{{\left| {{\mathbf{g}}_m^H{{\mathbf{w}}_k}} \right|}^2}} \right\}}}{{{\mathbb{E}}\left\{ {{{\left| {{\mathbf{g}}_m^H{{\mathbf{w}}_c}} \right|}^2}} \right\} + {\mathbb{E}}\left\{ {\sum\limits_{i = 1,i \ne k}^K {{{\left| {{\mathbf{g}}_m^H{{\mathbf{w}}_i}} \right|}^2}} } \right\} + {\mathbb{E}}\left\{ {{{\left| {{\mathbf{g}}_m^H{{\mathbf{w}}_v}} \right|}^2}} \right\} + \sigma _m^2}} \nonumber\\
		&= \frac{{{\mathbb{E}}\left\{ {{\mathbf{g}}_m^H{{\mathbf{W}}_k}{{\mathbf{g}}_m}} \right\}}}{{{\mathbb{E}}\left\{ {{\mathbf{g}}_m^H{{\mathbf{W}}_c}{{\mathbf{g}}_m}} \right\} + {\mathbb{E}}\left\{ {\sum\limits_{i = 1,i \ne k}^K {{\mathbf{g}}_m^H{{\mathbf{W}}_i}{{\mathbf{g}}_m}} } \right\} + {\mathbb{E}}\left\{ {{\mathbf{g}}_m^H{{\mathbf{W}}_v}{{\mathbf{g}}_m}} \right\} + \sigma _m^2}} \nonumber\\
		&= \frac{{{\rm{tr}}\left\{ {{{\mathbf{G}}_m}{{\mathbf{W}}_k}} \right\}}}{{{\rm{tr}}\left\{ {{{\mathbf{G}}_m}{{\mathbf{W}}_c}} \right\} + {\rm{tr}}\left\{ {{{\mathbf{G}}_m}\sum\limits_{i = 1,i \ne k}^K {{{\mathbf{W}}_i}} } \right\} + {\rm{tr}}\left\{ {{{\mathbf{G}}_m}{{\mathbf{W}}_v}} \right\} + \sigma _m^2}}
		\label{SNREpkm}
	\end{align}		
	\hrulefill
\end{figure*}

\color{black}

\subsection{Sensing Model}	

Depending on which signal is utilized to perform the sensing function, the sensing signal is expressed as
\begin{align}
	{\mathbf{X}} = {\alpha _1}{{\mathbf{w}}_c}{{{s}}_c} + {\alpha _2}{{\mathbf{w}}_{v}}{{{s}}_v},
\end{align}
where 
$\alpha _1$ and $\alpha _2$ are binary variables.
{Depending on the values of $\alpha _1$ and $\alpha _2$, the following schemes are considered:
\begin{enumerate}
	\item Scheme 1: ${\alpha _1} = 0$ and ${\alpha _2} = 1$. In this scheme, the extra signal is utilized to sense the targets. %
	
	\item Scheme 2: ${\alpha _1} = 1$ and ${\alpha _2} = 0$. In this scheme, the common stream is utilized for target sensing and the extra signal that is utilized as AN to improve the secrecy performance of the RSMA systems.
	\item Scheme 3: ${\alpha _1} = 1$ and ${\alpha _2} = 1$. Both the common stream and extra sequence are utilized for target sensing\footnote{
		A common point between Scheme 2 and Scheme 3 is that the common stream is utilized to sense the targets. The difference is whether the extra signal is used to sense. 
		This work considers the communication-centered scenarios while taking the sensing requirement as a constraint. Thus, Scheme 2 and Scheme 3 have similar performance, which is testified in \cite{XuC2021JSTSP} and Fig. \ref{fig02} of Section V. 		
	}. 
\end{enumerate}}

The received echo signal at $S$ is expressed as\footnote{
Like \cite{SuN2024TWC}, \cite{ZhaoB2025WCL}, since the radar BS generally has a strong signal separation capability, and a variety of methods can be used for filtering. We assume that $S$ can perfectly separate the echo from $T_t$ from the interference. 
}
\begin{align}
	{\mathbf{Y}}\left( {\bm{\theta }} \right) = {{\mathbf{A}}_r}\left( {\bm{\theta }} \right){\mathbf{BA}}_t^H\left( {\bm{\theta }} \right){\mathbf{X}} + {\mathbf{Q}},
\end{align}
where 
${{\mathbf{A}}_r} = \left[ {{{\mathbf{a}}_r}\left( {{\theta _1}} \right),...,{{\mathbf{a}}_r}\left( {{\theta _T}} \right)} \right]$, 
${{\mathbf{A}}_t} = \left[ {{{\mathbf{a}}_t}\left( {{\theta _1}} \right),...,{{\mathbf{a}}_t}\left( {{\theta _T}} \right)} \right]$, ${\mathbf{B}} = {\rm diag}\left( {{\beta _1},...,{\beta _T}} \right),{\bm{\theta }} = \left[ {{{\mathbf{\theta }}_1},...,{{\mathbf{\theta }}_T}} \right]$,  
${{\mathbf{a}}_r}\left( \theta  \right) = \left[ {1,{e^{j\frac{{2\pi }}{\lambda }d\sin \left( \theta  \right)}},...,{e^{j\frac{{2\pi }}{\lambda }d\left( {{N_r} - 1} \right)\sin \left( \theta  \right)}}} \right]$, 
$\theta_t$ is the angle of $T_t$, 
$\beta _t = b_{R_t} + jb_{I_t}$ denotes the target reflection complex amplitude, which is a normalization factor that depends on the radar cross section and path loss of the $t$-th target \cite{HuaH2024TWC, LiJ2008TSP, WuG2024IOT},
and 
${\bm{Q} }\sim CN\left( {0,{\sigma ^2}{\mathbf{I}}} \right)$ denotes the AWGN matrix. 

In this work, the aim of sensing is to estimate ${\bm\zeta} = [\bm{\theta},\mathbf{B}_R,\mathbf{B}_I]$, where 
${{\mathbf{B}}_R} = \left[ {{b_{{R_1}}}, \cdots ,{b_{{R_T}}}} \right]$ 
and 
${{\mathbf{B}}_I} = \left[ {{b_{{I_1}}}, \cdots ,{b_{{I_T}}}} \right]$.
The CRB of ${\bm\zeta}$ is utilized as the sensing performance metric and expressed as \cite{LiJ2008TSP}
\begin{align}
	\bm{\phi(\bm\zeta)} = {\mathbf{F}^{ - 1}},
\end{align}
where 
${\mathbf{F}}$ denotes the FIM, which is expressed as
\begin{align}
	\mathbf{F} = \frac{2}{\sigma ^2}\left[ {\begin{array}{*{20}{c}}
			{{\mathop{\rm Re}\nolimits} \left( {{{\mathbf{F }}_{11}}} \right)}&{{\mathop{\rm Re}\nolimits} \left( {{{\mathbf{F }}_{12}}} \right)}&{ - {\mathop{\rm Im}\nolimits} \left( {{{\mathbf{F }}_{12}}} \right)}\\
			{{{{\mathop{\rm Re}\nolimits} }^T}\left( {{{\mathbf{F }}_{12}}} \right)}&{{\mathop{\rm Re}\nolimits} \left( {{{\mathbf{F }}_{22}}} \right)}&{ - {\mathop{\rm Im}\nolimits} \left( {{{\mathbf{F }}_{22}}} \right)}\\
			{ - {{{\mathop{\rm Im}\nolimits} }^T}\left( {{{\mathbf{F }}_{12}}} \right)}&{ - {{{\mathop{\rm Im}\nolimits} }^T}\left( {{{\mathbf{F }}_{22}}} \right)}&{{\mathop{\rm Re}\nolimits} \left( {{{\mathbf{F }}_{22}}} \right)}
	\end{array}} \right],
\end{align}
where 
${{\mathbf{F}}_{11}} = L\left( {\dot {\mathbf{A }}_r^H{{\dot {\mathbf{A }}}_r}} \right) \odot \left( {{{\mathbf{B }}^*}{\mathbf{A }}_t^H{{\mathbf{R }}^*}{{\mathbf{A }}_t}{\mathbf{B }}} \right)+ L\left( {\dot {\mathbf{A }}_r^H{{\mathbf{A }}_r}} \right) \odot \left( {{{\mathbf{B }}^*}{\mathbf{A }}_t^H{{\mathbf{R }}^*}{{\dot {\mathbf{A }}}_t}{{\mathbf{B }}^T}} \right) + L\left( {{\mathbf{A }}_r^H{{\dot {\mathbf{A }}}_r}} \right) \odot \left( {{{\mathbf{B }}^*}\dot {\mathbf{A }}_t^H{{\mathbf{R }}^*}{{\mathbf{A }}_t}{{\mathbf{B }}^T}} \right) + L\left( {{\mathbf{A }}_r^H{{\mathbf{A }}_r}} \right) \odot \left( {{{\mathbf{B }}^*}\dot {\mathbf{A }}_t^H{{\mathbf{R }}^*}{{\dot {\mathbf{A }}}_t}{{\mathbf{B }}^T}} \right)$,
${{\mathbf{F}}_{12}} =L\left( {\dot {\mathbf{A }}_r^H{{\mathbf{A }}_r}} \right) \odot \left( {{{\mathbf{B }}^*}{\mathbf{A }}_t^H{{\mathbf{R }}^*}{{\mathbf{A }}_t}} \right) + L\left( {{\mathbf{A }}_r^H{{\mathbf{A }}_r}} \right) \odot \left( {{{\mathbf{B }}^*}\dot {\mathbf{A }}_t^H{{\mathbf{R }}^*}{{\mathbf{A }}_t}} \right)$, 
and ${{\mathbf{F}}_{22}} = L\left( {{\mathbf{A }}_r^H{{\mathbf{A }}_r}} \right) \odot \left( {{\mathbf{A }}_t^H{{\mathbf{R }}^*}{{\mathbf{A }}_t}} \right)$,
${\dot {\mathbf A}_r} = \left[ {\frac{{\partial {{\mathbf{a }}_r}}}{{\partial {\theta _1}}},...\frac{{\partial {{\mathbf{a }}_r}}}{{\partial {\theta _t}}}} \right]$
${\dot {\mathbf A}_t} = \left[ {\frac{{\partial {{\mathbf{a }}_t}}}{{\partial {\theta _1}}},...\frac{{\partial {{\mathbf{a }}_t}}}{{\partial {\theta _t}}}} \right]$,
$\mathbf{R}$ is the covariance matrix of the sensing signal $\mathbf X$ and is denoted by
\begin{align}
	{\mathbf{R}} ={\alpha _1} {\mathbf{W}}_c + {\alpha _2} {{{{\mathbf{W}}_v}}}.
\end{align}

\section{The Scenarios with Eavesdropping CSI}
\label{sec:WithECSI}

{In this section, the scenarios wherein the eavesdropping CSI is available are considered. 
	Two optimization problems are formulated to design the precoding matrices. 
	One is maximizing the minimum URPR by jointly designing the BF vectors and CRA subject to the QoS, security, and sensing constraints, and 
	the other is maximizing the minimum USRPR by jointly designing the beamforming vectors and secrecy CRA subject to the QoS and sensing constraints.  }

\subsection{Maximizing the Minimum User's Rate to BS's Power Ratio}

{
In this subsection, it is assumed that the confidential messages are transmitted in the private stream and the common stream is utilized as AN. Thus the achievable secrecy rate of $U_k$ is equal to
${\left[ {{R_{p,k}} - \mathop {\max }\limits_m {R_{p,k,m}^{\rm{E}}}} \right]^ + }$,
where 
${\left[ x \right]^ + } = \max \left( {x,0} \right)$.
To ensure that the common stream can not be decoded by all the eavesdroppers, the following constraint must be satisfied { \cite{MaoY2022Survey, YaoB2024ARX, LiuZ2024IOT, GaoP2023JSAC}} 
\begin{align}
	\mathop {\max }\limits_{1 \le m \le M} {R_{c,m}^{\rm{E}}} \le {R_c}.
\end{align}	}

We consider the problem of minimizing the transmit power at $S$ and maximizing the minimum total rate among all the LUs by constraining the QoS, security, and sensing performance. The following optimization problem is formulated
\begin{subequations}
	\begin{align}	
		\mathcal{P}_{1} \,:\, &\mathop {\max }\limits_{{\mathbf{W}}, {\mathbf{C}}} \mathop {\min }\limits_k \,\, {\eta _k}             \label{P1a}\\
		{\mathrm{s.t.}}\; 
		&{c_k} \ge 0, \forall k,         																	\label{P1b}\\
		&\sum\limits_{i = 1}^K {{c_i}}  \le {R_{c}},  									  		\label{P1c}\\	
		&{{{R_c} - \mathop {\max }\limits_{1 \le m \le M} R_{c,m}^{\rm{E}}} > 0,}       			\label{P1d}  \\	
		&{ {R_{p,k}} - \mathop {\max }\limits_{1 \le m \le M} R_{p,k,m}^{\rm{E}} \ge R_{\sec }^{\rm th}, \forall k, }\label{P1e}\\
		& {R_k^{{\rm{total}}}} \ge {R_{\rm{U}}^{{\rm{th}}}}, \forall k									\label{P1f}			\\	
		&\left| {{\bm \phi} \left( {\bm \varsigma } \right)} \right| \le \vartheta,   			\label{P1g}\\
		&P \le {P_{\max }},                 													\label{P1h}\\
		&{{\mathbf{W}}_q} \succeq 0, q \in \left\{ {c,k,v} \right\},        						\label{P1i}\\
		&{\mathrm {rank}}\left( {{{\mathbf{W}}_q}} \right) = 1, q \in \left\{ {c,k,v} \right\},		\label{P1j} 
	\end{align} 
\end{subequations}	
where {
${\eta _k} = \frac{{R_k^{{\rm{total}}}}}{P}$,
which denotes the URPR of $U_k$ in this work\footnote{ 	
	It is important to note that URPR of $U_k$ represents the ratio of $U_k$'s total rate to the transmit power of the ISAC BS. This metric differs from both user' EE and system's EE. In uplink scenarios, users' EE is defined as the ratio of the user’s achievable rate to its own transmit power. In downlink scenarios, system's EE is defined as the ratio of the sum rate to the BS’s transmit power.
}, }
${\mathbf{W}} = \left\{ {{{\mathbf{W}}_c},{{\mathbf{W}}_k},{{\mathbf{W}}_v}} \right\}$, 
$P = {\rm{tr}}\left( {{{\mathbf{W}}_c}} \right) + \sum\limits_{k = 1}^K {{\rm{tr}}\left( {{{\mathbf{W}}_k}} \right)}  + {\rm{tr}}\left( {{{{\mathbf{W}}_v}}} \right)$, 
${\mathbf{C}} = \left\{ {{c_k}, k = 1, 2, ..., K} \right\}$, 
and 
$\vartheta$ denotes the CRB requirement.
(\ref{P1b}) and (\ref{P1c}) are the common stream constraints for the LUs, 
(\ref{P1d}) ensures that the common stream can not be decoded by the eavesdroppers, 
(\ref{P1e}) denotes the secrecy constraints, 
(\ref{P1f}) denotes the total rate constraints, 
(\ref{P1g}) is the CRB constraint on the accuracy of the estimation of the target parameters, 
(\ref{P1h}) denotes the total power budget constraint of $S$, 
and 
(\ref{P1i}) and (\ref{P1j}) are positive semi-definite and rank one constraints.
{Due to the non-convexity of the fractional objective function (\ref{P1a}) and non-convex constraints (\ref{P1c})-(\ref{P1f}) and (\ref{P1j}), solving $\mathcal{P}_{1}$ is challenging.}

Firstly, utilizing the Dinkelbach’s method, $\mathcal{P}_{1}$ is reformulated as  
\begin{subequations}
	\begin{align}	
		\mathcal{P}_{1.1} &\mathop {\max }\limits_{{\mathbf{W}}, {\mathbf{C}},\lambda_1 } f\left( {\mathbf{W}},{\mathbf{C}} \right)  - \lambda_1 P   \label{P1.1a}\\
		&{\mathrm{s.t.}}\; {\rm (\ref{P1b})} - {\rm (\ref{P1j})},  \nonumber
	\end{align} 
\end{subequations}	
where 
$f\left( {\mathbf{W}}, {\mathbf{C}} \right) = \min \left( R_k^{\rm{total}} \right) = \min \left( {{c_k} + R_{p,k}} \right)$ 
and 
$\lambda_1$ is the penalty factor, which is obtained iteratively as follows 
\begin{align}	
	\lambda _1^{\left( {j + 1} \right)} = \frac{{{{\left( {\mathop {\min }\limits_k R_k^{\rm{total}}} \right)}^{\left( j \right)}}}}{{{P^{\left( j \right)}}}},
\end{align} 
where 
${\left(  \right)} ^{\left( j \right)}$ denotes the $j$-th iteration value.
 
Secondly, (\ref{P1c}) and (\ref{P1d}) are non-convex because the concavity of $R_{c,k}$ is uncertain. To deal this problem, 
based on (\ref{barSNRck}) and (\ref{Rck}),
$R_{c,k}$ is rewritten as
\begin{align}
	{R_{c,k}} & = \underbrace {{{\log }_2}\left( {{\rm{tr}}\left\{ {{{\mathbf{H}}_k}{{\mathbf{W}}_c}} \right\} + {\chi_1} + \sigma _k^2} \right)}_{ \buildrel \Delta \over = R_{c,k}^1} \nonumber \\
	& - \underbrace {{{\log }_2}\left( {{\chi_1} + \sigma _k^2} \right)}_{R_{c,k}^2}, \label{Rck22}
\end{align}
where 
${\chi_1} = {\rm{tr}}\left\{ {{{\mathbf{H}}_k}\sum\limits_{i = 1}^K {{{\mathbf{W}}_i}} } \right\} + {\rm{tr}}\left\{ {{{\mathbf{H}}_k}{{\mathbf{W}}_v}} \right\}$. 
It should be noted that both $R_{c,k}^1$ and $R_{c,k}^2$ are concave functions with respect to ${\mathbf W}$, which is proved in Appendix \ref{sec:appendicesA}. 
Then, the upper bound of $R_{c,k}^2$ is obtained as
\begin{align}
	R_{c,k}^2 &\le \frac{1}{{\ln 2}} \left({ {{{\left( {A_1^{\left( j \right)}} \right)}^{ - 1}} {{\chi_1}} }  - \left( {{{ {A_1^{\left( j \right)}} }^{ - 1}}\chi_1^{\left( j \right)}} \right)} \right) +{\log _2}\left(A_1^{\left( j \right)}\right) \nonumber\\
	& \buildrel \Delta \over = R_{c,k}^{2,{\rm {ub}}}, \label{Rck2up}
\end{align}
where 
$\chi_1^{\left( j \right)} = {\rm{tr}}\left\{ {{{\mathbf{H}}_k}\sum\limits_{i = 1}^K {{\mathbf{W}}_i^{\left( j \right)}} } \right\} + {\rm{tr}}\left\{ {{{\mathbf{H}}_k}{\mathbf{W}}_v^{\left( j \right)}} \right\}$ 
and 
$A_1^{\left( j \right)} = \chi_1^{\left( j \right)} + \sigma _k^2$.
Then, (\ref{P1c}) is reformulated as
\begin{align}
\sum\limits_{k = 1}^K {{c_k}}  \le R_{c,k}^1 - R_{c,k}^{2, {\rm{ub}}}, \forall k.	\label{p1.2c}	
\end{align}

Similarly, ${R_{c,m}^{\rm{E}}}$ is rewritten as
\begin{align}
	{R_{c,m}^{\rm{E}}} &= \underbrace {{{\log }_2}\left( {{\chi_2}} \right.\left. { + \sigma _m^2} \right)}_{ \buildrel \Delta \over = R_{c,m}^{{\rm{E}},1}} \nonumber \\
	&- \underbrace {{{\log }_2}\left( {{\rm{tr}}\left\{ {{{\mathbf{G}}_m}\sum\limits_{i = 1}^K {{{\mathbf{W}}_i}} } \right\} + {\rm{tr}}\left\{ {{{\mathbf{G}}_m}{{\mathbf{W}}_v}} \right\} + \sigma _m^2} \right)}_{ \buildrel \Delta \over = R_{c,m}^{{\rm{E}},2}},
\end{align}
where 
${\chi_2} = {\rm{tr}}\left\{ {{{\mathbf{G}}_m}{{\mathbf{W}}_c}} \right\} + {\rm{tr}}\left\{ {{{\mathbf{G}}_m}\sum\limits_{i = 1}^K {{{\mathbf{W}}_i}} } \right\} + {\rm{tr}}\left\{ {{{\mathbf{G}}_m}{{\mathbf{W}}_v}} \right\}$.
By utilizing the SCA technique, we have 
\begin{align}
	R_{c,m}^{{\rm{E}},1} &\le \frac{1}{{\ln 2}} \left({ {{{\left( {A_2^{\left( j \right)}} \right)}^{ - 1}}{{\chi_2}} }  - \left( {{{{A_2^{\left( j \right)}} }^{ - 1}}\chi_2^{\left( j \right)}} \right)} \right) +{\log _2}\left(A_2^{\left( j \right)}\right) \nonumber\\
	&\buildrel \Delta \over = R_{c,m}^{{\rm{E}}, 1, {\rm{ub}}},
\end{align}
where 
$\chi_2^{\left( j \right)} = {\rm{tr}}\left\{ {{{\mathbf{G}}_m}{\mathbf{W}}_c^{\left( j \right)}} \right\} + {\rm{tr}}\left\{ {{{\mathbf{G}}_m}\sum\limits_{i = 1}^K {{\mathbf{W}}_i^{\left( j \right)}} } \right\} + {\rm{tr}}\left\{ {{{\mathbf{G}}_m}{\mathbf{W}}_v^{\left( j \right)}} \right\}$ and $A_2^{\left( j \right)} = \chi_2^{\left( j \right)} + \sigma _m^2$.
Then, (\ref{P1d}) is reformulated as
\begin{align}
	R_{c,k}^1 - R_{c,k}^{2, {\rm{ub}}} \ge R_{c,m}^{{\rm{E}}, 1, {\rm{ub}}} - R_{c,m}^{{\rm{E}},2}, \forall k, m.	\label{p1.2d}	
\end{align}
	
{(\ref{P1e}) and (\ref{P1f}) are non-convex because the concavity of ${R_{p,k}}$ and $R_{p,k,m}^{\rm{E}}$ is uncertain. }
With the same method, ${R_{p,k}}$ is rewritten as
\begin{align}	
	{R_{p,k}} &= \underbrace {{{\log }_2}\left( {{\rm{tr}}\left\{ {{{\mathbf{H}}_k}{{\mathbf{W}}_k}} \right\} + {\chi_3} + \sigma _k^2} \right)}_{ \buildrel \Delta \over = R_{p,k}^1} \nonumber\\
	&- \underbrace {{{\log }_2}\left( {{\chi_3} + \sigma _k^2} \right)}_{ \buildrel \Delta \over = R_{p,k}^2},
\end{align}
where  
${\chi_3} = {\rm{tr}}\left\{ {{{\mathbf{H}}_k}\sum\limits_{i = 1,i \ne k}^K {{{\mathbf{W}}_i}} } \right\} + {\rm{tr}}\left\{ {{{\mathbf{H}}_k}{{\mathbf{W}}_v}} \right\}$. 
It should be noted that both $R_{p,k}^1$ and $R_{p,k}^2$ are concave functions with respect to ${\mathbf W}$.
Similar to (\ref{Rck2up}), the upper bound of the convex approximation to $R_{p,k}^2 $ is obtained as
\begin{align}
	R_{p,k}^2 & \le  \frac{1}{{\ln 2}}\left( { {{{\left( {A_3^{\left( j \right)}} \right)}^{ - 1}}{\chi_3}} - {{{\left( {A_3^{\left( j \right)}} \right)}^{ - 1}}\chi_3^{\left( j \right)}} } \right) +{\log _2}\left(A_3^{\left( j \right)}\right) \nonumber\\
	& \buildrel \Delta \over = R_{p,k}^{2, {\rm{ub}}},
\end{align}
where 
$\chi_3^{\left( j \right)} = {\rm{tr}}\left\{ {{{\mathbf{H}}_k}\sum\limits_{i = 1,i \ne k}^K {{\mathbf{W}}_i^{\left( j \right)}} } \right\} + {\rm{tr}}\left\{ {{{\mathbf{H}}_k}{\mathbf{W}}_v^{\left( j \right)}} \right\}$ 
and 
$A_3^{\left( j \right)} = \chi_3^{\left( j \right)} + \sigma _k^2$.

With the same method as (\ref{Rck22}), $R_{p,k,m}^{\rm{E}}$ is rewritten as
\begin{align}
	R_{p,k,m}^{\rm{E}} &= \underbrace {{{\log }_2}\left( {{\rm{tr}}\left\{ {{{\mathbf{G}}_m}{{\mathbf{W}}_k}} \right\} + {\chi_4} + \sigma _m^2} \right)}_{ \buildrel \Delta \over = R_{p,k,m}^{{\rm{E}},1}} \nonumber\\
	& - \underbrace {  {{\log }_2}\left( {{\chi_4} + \sigma _m^2} \right)}_{ \buildrel \Delta \over = R_{p,k,m}^{{\rm{E}},2}},
\end{align}
where 
${\chi_4} = {\rm{tr}}\left\{ {{{\mathbf{G}}_m}{{\mathbf{W}}_c}} \right\} + {\rm{tr}}\left\{ {{{\mathbf{G}}_m}\sum\limits_{i = 1,i \ne k}^K {{{\mathbf{W}}_i}} } \right\} + {\rm{tr}}\left\{ {{{\mathbf{G}}_m}{{\mathbf{W}}_v}} \right\}$.
Then, the lower bound of the ${R_{p,k,m}^{{\rm{E}},1}}$ is obtained as
\begin{align}
	{R_{p,k,m}^{{\rm{E}},1}} &\le \frac{1}{{\ln 2}} \left({ {{{\left( {A_4^{\left( j \right)}} \right)}^{ - 1}} {{\chi_4}} }  - {{{\left( {A_4^{\left( j \right)}} \right)}^{ - 1}}\chi_4^{\left( j \right)}} } \right) +{\log _2}\left(A_4^{\left( j \right)}\right) \nonumber\\
	& \buildrel \Delta \over = R_{p,k,m}^{{\rm{E}}, 1, {\rm{ub}}},	
\end{align}
where 
$\chi_4^{\left( j \right)} = {\rm{tr}}\left\{ {{{\mathbf{G}}_m}{\mathbf{W}}_c^{\left( j \right)}} \right\} + {\rm{tr}}\left\{ {{{\mathbf{G}}_m}\sum\limits_{i = 1,i \ne k}^K {{\mathbf{W}}_i^{\left( j \right)}} } \right\} + {\rm{tr}}\left\{ {{{\mathbf{G}}_m}{\mathbf{W}}_v^{\left( j \right)}} \right\}$ 
and
$A_4^{\left( j \right)} = {\rm{tr}}\left\{ {{{\mathbf{G}}_m}{\mathbf{W}}_k^{\left( j \right)}} \right\} + \chi_4^{\left( j \right)}$.
Then, (\ref{P1e}) and (\ref{P1f}) are reformulated as 
\begin{align}
	&R_{p,k}^1 - R_{p,k}^{2, {\rm{ub}}} - (R_{p,k,m}^{{\rm{E}}, 1, {\rm{ub}}} - R_{p,k,m}^{{\rm{E}},2}) \ge R_{\sec }^{\rm th}, \forall k, m	\label{p1.2e}
\end{align}
and 
\begin{align}
	c_k+R_{p,k}^1 - R_{p,k}^{2, {\rm{ub}}}  \ge {R_{\rm{U}}^{{\rm{th}}}}, \forall k,	\label{p1.2f}
\end{align}
respectively.

Finally, by introducing the slack variable $t$, $\mathcal{P}_{1.1}$ is rewritten as
\begin{subequations}
	\begin{align}
		\mathcal{P}_{1.2} \,:\, & \mathop {\max }\limits_{{\mathbf{W}}, {\mathbf{C}},t} t\\
		{\mathrm{s.t.}}\; 
		&\hat f\left( {\mathbf{W}}, {\mathbf{C}} \right) - \lambda_1 P  \ge t,    \label{p1.2b}		\\
		&({\rm \ref{P1b}}),({\rm \ref{P1g}})-({\rm \ref{P1j}})		\nonumber			\\
		& ({\rm \ref{p1.2c}}), ({\rm \ref{p1.2d}}), ({\rm \ref{p1.2e}}), ({\rm \ref{p1.2f}}), 			\nonumber	
	\end{align} 
\end{subequations}	
where $\hat f\left( {\mathbf{W}}, {\mathbf{C}} \right) = \min \left( {{c_k} + \left( {R_{p,k}^1 - R_{p,k}^{2, {\rm{ub}}}} \right)} \right) $.
By ignoring the Rank-1 constraint (\ref{P1j}), $\mathcal{P}_{1.2}$ is a convex problem that can be solved to obtain ${\mathbf{W}}$.

For (\ref{P1j}), the rank-1 of ${\mathbf{W}}_q$ means that the matrix ${\mathbf{W}}_q$  has only one non-zero eigenvalue, which is equivalent to the fact that the trace of ${\mathbf{W}}_q$ is equal to its largest eigenvalue, which is expressed as \cite{LiuZ2024TWC}
\begin{align}
	{\mathrm{rank}}\left( {{{\mathbf{W}}_q}} \right) = 1 \Leftrightarrow  {\mathrm{tr}}\left( {{{\mathbf{W}}_q}} \right) = {\chi _{{q}}},
	\label{Eq41}
\end{align}
where $\chi _{q}$ denotes the maximum eigenvalue obtained from the eigen decomposition of ${{\mathbf{W}}_q}$.   
(\ref{Eq41}) is non-convex and the following iterative algorithm is utilized  
\begin{align}
	{\mathrm{tr}}\left( {{{\mathbf{W}}_q}} \right) = {\chi _q} \Rightarrow {\mathrm{tr}}\left( {{{\mathbf{W}}_q}} \right) = {\left( {{\bf{u}}_q^{\left( {j } \right)}} \right)^H}{{\mathbf{W}}_q}{\bf{u}}_q^{\left( {j } \right)},
	\label{Eq46}
\end{align}
where 
${{\bf{u}}_q}$ is the eigenvector corresponding to the largest eigenvalue $\chi _q$ of ${{\mathbf{W}}_q}$. 
Then, by utilizing a penalty function in the objective function, the optimization problem is expressed as \cite{LiuZ2024TWC}
\begin{subequations}
	\begin{align}
		\mathcal{P}_{1.3} & \mathop {\max }\limits_{{\mathbf{W}}, {\mathbf{C}},t} t - {\rho _1}\left( { {\mathrm{tr}}\left( {{{\mathbf{W}}_c}} \right) - {{\left( {{\bf{u}}_c^{\left( {j } \right)}} \right)}^H}{{\mathbf{W}}_c}{\bf{u}}_c^{\left( {j } \right)}} \right) \nonumber\\ 
		& \qquad - {\rho _1}\sum\limits_{k = 1}^K {\left( { {\mathrm{tr}}\left( {{{\mathbf{W}}_k}} \right) - {{\left( {{\bf{u}}_k^{\left( {j } \right)}} \right)}^H}{{\mathbf{W}}_k}{\bf{u}}_k^{\left( {j } \right)}} \right)} \nonumber\\ 
		&  \qquad- {\rho _1}\left( { {\mathrm{tr}}\left( {{{{\mathbf{W}}_v}}} \right) - {{\left( {{\bf{u}}_v^{\left( {j } \right)}} \right)}^H}{{{{\mathbf{W}}_v}}}{\bf{u}}_v^{\left( {j } \right)}} \right)\\
		{\mathrm{s.t.}}\; 
		& ({\rm \ref{P1b}}), ({\rm \ref{P1g}})-({\rm \ref{P1i}}),  										\nonumber	\\
		& ({\rm \ref{p1.2c}}), ({\rm \ref{p1.2d}}), ({\rm \ref{p1.2e}}), ({\rm \ref{p1.2f}}), ({\rm \ref{p1.2b}}),		\nonumber			
	\end{align} 
\end{subequations}	
where $\rho _1$ denotes the penalty factor.

$\mathcal{P}_{1.3}$ is a standard convex problem that can be solved using the CVX toolbox. The algorithm is summarized as \textbf{Algorithm 1}, where $j \le J_{\max}$ denotes the maximum number of iterations.
	
\begin{algorithm}[t]
\caption{Iterative Procedure of $\mathcal{P}_{1.3}$}
\KwIn{Initialize ${P_{\max }}$, $R_{\rm U}^{\rm th}$, $R_{\sec }^{\rm th}$, $\rho_1$, $\tau$, $\vartheta$, ${\mathbf{W}}_q^{\left( 0 \right)}$, ${\bf{u}}_q^{\left( 0 \right)}$, $\lambda_1^{\left( 0 \right)}$, $j \leftarrow 0$}
\Do{$\| opt^{(j+1)}-opt^{(j)}\| > \tau$ {\rm or} $j < J_{\max}$}
{   1. Solve $\mathcal{P}_{1.3}$ and obtain ${\mathbf W}_q^{\left( j+1 \right)}$;\\
	2. Obtain ${{{\bf{u}}_q}^{\left( j + 1 \right)}}$ by decomposing  ${\mathbf{W}}_q^{\left( j+1 \right)}$;\\     
	3. $\lambda _1^{\left( {j + 1} \right)} = \frac{{{{\left( {\mathop {\min }\limits_k R_k^{\rm{total}}} \right)}^{\left( j + 1 \right)}}}}{{{P^{\left( j + 1 \right)}}}}$;\\
	4. $j = j + 1$;
}
\KwOut{${\mathbf W}_q^{\left( j \right)}$}

\end{algorithm}

\begin{algorithm}[t]
	\caption{Iterative Procedure of $\mathcal{P}_{2.1}$}
	\KwIn{Initialize ${P_{\max }}$, $R_{\sec }^{\rm th}$, $\rho_2$, $\tau$, $\vartheta$,
		${\mathbf{W}}_q^{\left( 0 \right)}$, ${\bf{u}}_q^{\left( 0 \right)}$, $\lambda_2^{\left( 0 \right)}$, $j \leftarrow 0$.}
		\Do{$\| opt^{(j+1)}-opt^{(j)}\| > \tau$ {\rm or} $j < J_{\max}$}
		{   1. Solve $\mathcal{P}_{2.1}$ and obtain ${\mathbf W}_q^{\left( j+1 \right)}$;\\
			2. Obtain ${{{\bf{u}}_q}^{\left( j+1 \right)}}$ by decomposing  ${\mathbf{W}}_q^{\left( j+1 \right)}$;\\   
			3. $\lambda _2^{\left( {j + 1} \right)} = \frac{R_k^{\sec ,{\rm{lb,}}\left( j + 1 \right)}}{P^{\left( j + 1 \right)}}$;\\
			4. $j = j + 1$;	
		}
	\KwOut{${\mathbf W}_q^{\left( j \right)}$}
\end{algorithm}	
	
\subsection{Maximizing the Minimum User's Secrecy Rate to BS's Power Ratio}

{In this subsection, it is assumed that the confidential messages are transmitted in both common and private streams. 
We define $c_{c,k}^{\sec }$ as the non-negative common secrecy rate for $U_k$ is allocated. }
Then we have \cite{GaoY2023CL}
\begin{align}
	\sum\limits_{k = 1}^K {c_{c,k}^{\sec }} \le \left( {{R_c} - \mathop {\max }\limits_{1 \le m \le M} R_{c,m}^{\rm{E}}} \right).
	\label{commSR}
\end{align}

The achievable secrecy rate of $U_k$ is obtained as
\begin{align}
	R_k^{\sec } = c_{c,k}^{\sec } + {\left[ {{R_{p,k}} - \mathop {\max }\limits_{1 \le m \le M} {R_{p,k,m}^{\rm{E}}}} \right]^ + }.
\end{align}

{We consider the problem of minimizing the transmit power at $S$ and maximizing the minimum secrecy rate among all the LUs by constraining the security requirement and sensing constraint.} 
The following optimization problem is formulates
\begin{subequations}
	\begin{align}	
		\mathcal{P}_{2} \,:\, &\mathop {\max }\limits_{{\mathbf{W}}, {\mathbf{\tilde C}}} \mathop {\min }\limits_{k}\,\,\, {\eta _k^{\sec }}     \label{p02_a} \\
		{\mathrm{s.t.}}\;
		&	c_{c,k}^{\sec } \ge 0,  \forall k,                    				\label{p02_b} \\
		&{	\sum\limits_{k = 1}^K {c_{c,k}^{\sec }} \le \left( {{R_c} - \mathop {\max }\limits_{1 \le m \le M} R_{c,m}^{\rm{E}}} \right), }												\label{p02_c} \\
		&{	{R_{p,k}} - \mathop {\max }\limits_{1 \le m \le M} R_{p,k,m}^{\rm{E}} \ge 0, \forall k,} \label{p02_d}\\
		& {{{R_k^{\sec }}}} \ge R_{\sec }^{\rm th}, 	\forall k,						\label{p02_e}	\\
		& {\rm (\ref{P1g})} - {\rm (\ref{P1j})}, 								\nonumber
	\end{align}
\end{subequations}
where {
$\eta _k^{\sec } = \frac{{R_k^{\sec }}}{P}$ denotes the USRPR of $U_k$ and 
${\mathbf{\tilde C}} = \left\{ {c_{c,k}^{\sec }}, k = 1, 2, ..., K \right\}$. }
(\ref{p02_b})-(\ref{p02_c}) are secrecy constraint for the common stream 
and 
(\ref{p02_d}) is secrecy constraint for the private stream.
$\mathcal{P}_{2}$ is challenging to solve because of the non-convexity of (\ref{p02_b})-(\ref{p02_e}) and (\ref{P1g})-(\ref{P1j}).

Firstly, to deal with $R_k^{\sec }$ in (\ref{p02_e}), its lower bound is utilized and expressed as
\begin{align}
	R_k^{\sec } &\ge c_{c,k}^{\sec } + R_{p,k}^1 - R_{p,k}^{2, {\rm{ub}}} - {R_{p,k,m}^{{\rm{E}}, 1, {\rm{ub}}} + R_{p,k,m}^{{\rm{E}},2}} \nonumber \\
	&\buildrel \Delta \over= R_k^{\sec, {\rm {lb}}}.
\end{align}
By introducing the slack variable $t_2$, the optimization objective function of $\mathcal{P}_{2}$ is rewritten as
\begin{align}
	R_k^{\sec, {\rm {lb}}} - {\lambda _2}P \ge {t_2}, \forall k, \label{eq46}
\end{align}
where $\lambda _2$ is obtained as
\begin{align}	
	\lambda _2^{\left( {j + 1} \right)} = \frac{R_k^{\sec ,{\rm{lb,}}\left( j \right)}}{P^{\left( j \right)}}.
\end{align} 
Then, with the same method as $\mathcal{P}_{1}$, $ \mathcal{P}_{2}$ is relaxed as
\begin{subequations}
	\begin{align}
		\mathcal{P}_{2.1} \,:\, &\mathop {\max }\limits_{{\mathbf{W}}, {\mathbf{\tilde C}},t_2} t_2 - {\rho _2}\left( { {\mathrm{tr}}\left( {{{\mathbf{W}}_c}} \right) - {{\left( {{\bf{u}}_c^{\left( {j } \right)}} \right)}^H}{{\mathbf{W}}_c}u_c^{\left( {j } \right)}} \right) \nonumber\\ 
		& \qquad - {\rho _2}\sum\limits_{k = 1}^K {\left( { {\mathrm{tr}}\left( {{{\mathbf{W}}_k}} \right) - {{\left( {{\bf{u}}_k^{\left( {j } \right)}} \right)}^H}{{\mathbf{W}}_k}{\bf{u}}_k^{\left( {j } \right)}} \right)} \nonumber\\ 
		&  \qquad- {\rho _2}\left( { {\mathrm{tr}}\left( {{{{\mathbf{W}}_v}}} \right) - {{\left( {{\bf{u}}_v^{\left( {j } \right)}} \right)}^H}{{{{\mathbf{W}}_v}}}{\bf{u}}_v^{\left( {j } \right)}} \right)\\
		{\mathrm{s.t.}}\;
		&\sum\limits_{k = 1}^K {c_{c,k}^{\sec } \le \left( {R_{c,k}^1 - R_{c,k}^{2, {\rm{ub}}}} \right) - } \left( {R_{c,m}^{1, {\rm{ub}}} - R_{c,m}^2} \right), 	\forall k, m,				\\
		&\left( {R_{p,k}^1 - R_{p,k}^{2, {\rm{ub}}}} \right) - \left( {R_{p,k,m}^{{\rm{E}}, 1, {\rm{ub}}} - R_{p,k,m}^{{\rm{E}},2}} \right) \ge 0,	\forall k, m,									\\
		&R_k^{\sec, {\rm {lb}}} \ge R_{\sec }^{\rm th},	\forall k,	\\
		& {\rm (\ref{P1g})} - {\rm (\ref{P1i})}, ({\rm \ref{p02_b}}), ({\rm \ref{eq46}}), 		\nonumber
	\end{align}
\end{subequations}
where $\rho _2$ represents the penalty factor.

$ \mathcal{P}_{2.1}$ is a standard convex problem that can be solved by an iterative optimization algorithm, which is summarized in \textbf{Algorithm 2}.

\begin{algorithm}[t]
	\caption{Iterative Procedure of $\mathcal{P}_{3.1}$}
	\KwIn{ Initialize ${P_{\max }}$, $R_{\rm U}^{\rm th}$, $\rho_3$, $\tau$, $\vartheta$,
		${\mathbf{W}}_q^{\left( 0 \right)}$, ${\bf{u}}_q^{\left( 0 \right)}$, $j \leftarrow 0$.}
	\Do{$\| opt^{(j+1)}-opt^{(j)}\| > \tau$ {\rm or} $j < J_{\max}$}
	{
		1. Solve $\mathcal{P}_{3.1}$ and obtain ${\mathbf W}_q^{\left( j+1 \right)}$;\\
		2. Obtain ${{{\bf{u}}_q}^{\left( j+1 \right)}}$ by decomposing  ${\mathbf{W}}_q^{\left( j+1 \right)}$;\\    
		3. $j = j + 1$;	
	}
	\KwOut{${\mathbf W}_q^{\left( j \right)}$}
\end{algorithm}
		
\section{The Scenarios Without Eavesdropping CSI}
\label{sec:WithoutECSI}

When the eavesdropper's CSI is unknown, on the premise of ensuring the communication and sensing performance, we use all the remaining energy to send the omni-directional AN to improve the security of communication, like refs. \cite{ChuJJ2023TVT} and \cite{SuN2024TWC}.
The transmitted signal at $S$ is expressed as
\begin{align}
	{\mathbf{x}} = {{\mathbf{w}}_c}{{{s}}_c} + \sum\limits_{k \in K} {{{\mathbf{w}}_k}{{{s}}_k}}  + {{\mathbf{w}}_{v}}{{{s}}_v} + {{\mathbf{w}}_{\rm{AN}}}{{{s}}_{\rm{AN}}}
\end{align}	
where 
${{{s}}_{\rm{AN}}}$ is the AN symbol which is assumed to be zero mean and unit power and 
${{\mathbf{w}}_{\rm{AN}}}$ is the transmission vector with ${\rm{diag}}\left( {{{\mathbf{W}}_{\rm{AN}}}} \right) = \frac{{\mathbf{I}}}{{{N_t}}}\left( {{P_{\max }} - P} \right)$ 
and ${\mathbf{W}}_{\rm{AN}}={\mathbf w}_{\rm{AN}}{\mathbf w}_{\rm{AN}}^H$.
It is assumed that $U_k$ and $S$ can perfectly eliminate ${{{s}}_{\rm{AN}}}$, like \cite{LiuZ2024TWC}.
Consequently, the BF matrices are designed to minimize the NPC for communication and sensing, thereby improving the system's secrecy performance. \color{black}
The formulated problem is expressed as	
\begin{subequations}
	\begin{align}
		\mathcal{P}_{3} \,:\, &\mathop {\min }\limits_{{\mathbf{W}}, {\mathbf{C}}} P   			\label{p03_a}\\
		{\mathrm{s.t.}}\;
		&{c_k} \ge 0, \forall k,     															\label{p03_b}\\
		&\sum\limits_{i = 1}^K {{c_i}}  \le {R_{c,k}},  \forall k,						  		\label{p03_c}\\		
		&{ {R_k^{{\rm{total}}}} \ge {R_{\rm{U}}^{{\rm{th}}}}, \forall k,}			\label{p03_d}\\
		& {\rm (\ref{P1g})}, {\rm (\ref{P1i})}, {\rm (\ref{P1j})}. 		\nonumber
	\end{align}
\end{subequations}

Based on the results in Section III.A, $\mathcal{P}_{3}$ is rewritten as 
\begin{subequations}
	
	\begin{align}
		\mathcal{P}_{3.1} \,:\, &\mathop {\min }\limits_{{\mathbf{W}}, {\mathbf{C}}} {\mathrm{tr}}\left( {{{\mathbf{W}}_c}} \right) + \sum\limits_{k = 1}^K {{\mathrm{tr}}\left( {{{\mathbf{W}}_k}} \right)} + {\mathrm{tr}}\left( {{{{{{\mathbf{W}}_v}}}}} \right)  \nonumber\\ 
		& \qquad+ {\rho _3}\left( {{\mathrm{tr}}\left( {{{\mathbf{W}}_c}} \right) - {{\left( {{\bf{u}}_c^{\left( {j } \right)}} \right)}^H}{{\mathbf{W}}_c}{\bf{u}}_c^{\left( {j } \right)}} \right) \nonumber\\ 
		& \qquad+ {\rho _3}\sum\limits_{k = 1}^K {\left( {{\mathrm{tr}}\left( {{{\mathbf{W}}_k}} \right) - {{\left( {{\bf{u}}_k^{\left( {j } \right)}} \right)}^H}{{\mathbf{W}}_k}{\bf{u}}_k^{\left( {j } \right)}} \right)}\nonumber\\ 
		&  \qquad+ {\rho _3}\left( { {\mathrm{tr}}\left( {{{{\mathbf{W}}_v}}} \right) - {{\left( {{\bf{u}}_v^{\left( {j } \right)}} \right)}^H}{{{{\mathbf{W}}_v}}}{\bf{u}}_v^{\left( {j } \right)}} \right)\\
		{\mathrm{s.t.}}\;
		&\sum\limits_{k = 1}^K {{c_k}}  \le R_{c,k}^1 - R_{c,k}^{2, {\rm{ub}}}, \forall k,\\
		&{c_k} + R_{p,k}^1 - R_{p,k}^{2, {\rm{ub}}} \ge {R_{\rm{U}}^{{\rm{th}}}}, \forall k,\\
		& {\rm (\ref{P1g})}, {\rm (\ref{P1i})}, ({\rm \ref{p03_b}}),		\nonumber
	\end{align}
\end{subequations}
where $\rho _3$ represents the penalty factor.

$\mathcal{P}_{3.1}$ is a standard convex problem that can be solved iteratively using the CVX toolbox.  The relevant algorithm is summarized as \textbf{Algorithm 3}.

The algorithm complexity analysis is as follows. 
$\mathcal{P}_{1.2}$  and $\mathcal{P}_{2.1}$ are composed of $K+2$ linear matrix inequality (LMI) constraints of size $N_t$ and $2KM+4K+3$ LMI constraints of size 1.
With the given convergence accuracy $\tau$ and the number of iterations ${J_{\max}}$, the computational complexity of algorithm 1 and algorithm 2 are expressed as  $O\left( {J_{\max}}{\sqrt {2KM + K{N_t}} {K^3}N_t^6} \log \left( {\frac{1}{\tau }} \right)\right)$.
$\mathcal{P}_{3.1}$ is composed by $K + 2$ LMI constraints of size $N_t$, ${4K + {N_t} + 2}$ LMI constraints of size 1, the computational complexity is $O\left( {{{J_{\max}}}\left( {{K^{3.5}}N_t^{6.5}} \right)} \log \left( {\frac{1}{\tau }} \right)\right)$.

\begin{table}
	\centering
	
	\caption{{Simulation Parameters}}
	\begin{tabular}{c|c|c|c}
		\Xhline{1.2pt}
		Parameters    	&   Value            	&    Parameters    	&   Value      \\
		\hline
		${d}_{u_k}$ 	&  60, 80, 100 m			& ${d}_{m} $ 		&70, 90 m\\
		\hline
		$\theta _{k}	$ &$ -60^ \circ, -5^ \circ, 50^ \circ$& $\theta _{m}$	   & $-15^ \circ, 40^ \circ$ \\
		\hline
		$\theta _T $	& $-30^ \circ, 30^ \circ$ & $\beta _0$				& $-30$ dB	 			\\
		\hline
		$ \sigma_k^2=\sigma_m^2$ & $-70$ dBm 				&   $\alpha_u $              &  2.2     \\
		\hline
		$P_{\max}$							& 30 dBm & $\vartheta$				& $-70$ dB					\\ 	
		\hline
		$N_r = N_t $ 				&  12 			& $\rho _k = \rho _m$  		& 100				\\		
		\hline
		${R_{\mathrm{U}}^{{\mathrm{th}}}}$  		&3	bps/Hz 		& $R_{\sec }^{\mathrm th}$			&  1 bps/Hz			\\ 	
		\Xhline{1.2pt}
	\end{tabular}
	\label{table3}
\end{table}

\section{Numerical Results}
\label{sec:Simulation}
	
This section presents numerical results to validate the convergence and effectiveness of the proposed schemes. The detailed parameter configurations of the ISAC system are summarized in Table \ref{table3}. 
	
To verify the effectiveness of the proposed algorithms, the following schemes are considered as benchmarks, which are described as follows.
\begin{itemize}
	\item Benchmark 1: In this scheme, similar to Refs. \cite{XuC2021JSTSP} and \cite{YinY2022CL}, there is no extra signal and the common stream is utilized for target sensing (denoted as `Ben1')\footnote{
		This scheme can be seemed as the extension of \cite{XuC2021JSTSP} and \cite{YinY2022CL} to the scenarios that the security is required.}. 
		This scheme is realized part by setting ${{\mathbf{w}}_{v}} = 0$, ${\alpha _1} = 1$, and ${\alpha _2} = 0$. 	

	\item Benchmark 2: In this scheme, the SDMA scheme is utilized with extra sequence that is utilized for target sensing (denoted as `SDMA').  
	The corresponding optimization problems are expressed as 
\begin{subequations}
	\begin{align}	
		\mathcal{P}_{1,{\mathrm {SDMA}}} \,:\, &\mathop {\max }\limits_{{\mathbf{W}_k}, {\mathbf{W}_v}} {\mathop {\min }\limits_k } \;\; {\eta _{k,{\mathrm{SDMA}}}}       															\label{P1SDa}\\
		{\mathrm{s.t.}}\; 
		& {R_{k,{\mathrm{SDMA}}}} \ge {R_{\mathrm{U}}^{{\mathrm{th}}}}, \forall k, 							\label{P1SDb}\\
		&R_{k,{\mathrm{SDMA}}}^{\sec } \ge R_{\sec }^{\mathrm {th}}, \forall k,							\label{P1SDc}\\
		&{{\mathbf{W}}_q} \succeq 0, q \in \left\{ {k,v} \right\},        						\label{P1SDd}\\
		&{\mathrm {rank}}\left( {{{\mathbf{W}}_q}} \right) = 1, q \in \left\{ {k,v} \right\},	\label{P1SDe} \\
		&{P_{{\mathrm{SDMA}}}} \le {P_{\max }},																		\label{P1SDf} \\						
		&({\rm \ref{P1g}}), 		\nonumber			
	\end{align} 
\end{subequations}	
\begin{subequations}
	\begin{align}	
		\mathcal{P}_{2,{\mathrm {SDMA}}} \,:\, &\mathop {\max }\limits_{{\mathbf{W}_k}, {\mathbf{W}_v}} {\mathop {\min }\limits_k }\;\; \eta _{k,{\mathrm{SDMA}}}^{\sec }     									\label{P2SDa}\\
		{\rm{s.t.}}\; 
		& ({\rm \ref{P1SDc}})- ({\rm \ref{P1SDf}}),	({\rm \ref{P1g}}),	\nonumber			
	\end{align} 
\end{subequations}	
\begin{subequations}
	\begin{align}	
		\mathcal{P}_{3,{\mathrm {SDMA}}} \,:\, &\mathop {\min }\limits_{{\mathbf{W}}_k,{{\mathbf{W}}_v}}  {{P _{{\mathrm{SDMA}}}}}       												\label{P3SDa}\\
		{\rm{s.t.}}\; 
		& ({\rm \ref{P1SDb}}), ({\rm \ref{P1SDd}}), ({\rm \ref{P1SDe}}), ({\rm \ref{P1g}}),		\nonumber		
	\end{align} 
\end{subequations}	
where 
${\eta _{k,{\mathrm{SDMA}}}} = \frac{{{R_{k,{\mathrm{SDMA}}}}}}{{{P_{{\mathrm{SDMA}}}}}}$, 
${R_{k,{\mathrm{SDMA}}}} = {\log _2}\left( {1 + \frac{{{\mathrm{tr}}\left\{ {{{\mathbf{H}}_k}{{\mathbf{W}}_k}} \right\}}}{{{\mathrm{tr}}\left\{ {{{\mathbf{H}}_k}\sum\limits_{i \ne k} {{{\mathbf{W}}_i}} } \right\} + {\mathrm{tr}}\left\{ {{{\mathbf{H}}_k}{{\mathbf{W}}_v}} \right\} + \sigma _k^2}}} \right)$, 
${P_{{\mathrm{SDMA}}}} = \sum\limits_{k = 1}^K {{\mathrm {tr}}\left( {{{\mathbf{W}}_k}} \right)}  + {\mathrm {tr}}\left( {{\mathbf{W}}_v} \right)$, 
$\eta _{k,{\mathrm{SDMA}}}^{\sec } = \frac{{R_{k,{\mathrm{SDMA}}}^{\sec }}}{{{P_{{\mathrm{SDMA}}}}}}$, 
$R_{k,{\mathrm{SDMA}}}^{\sec } = {\left[ {{R_{k,{\mathrm{SDMA}}}} - \mathop {\max }\limits_{1 \le m \le M} R_{k,m,{\mathrm{SDMA}}}^{\mathrm{E}}} \right]^ + }$, 	
and 
$R_{k,m,{\mathrm{SDMA}}}^{\mathrm{E}} = {\log _2}\left( {1 + \frac{{{\mathrm{tr}}\left\{ {{{\mathbf{G}}_m}{{\mathbf{W}}_k}} \right\}}}{{{\mathrm{tr}}\left\{ {{{\mathbf{G}}_m}\sum\limits_{j \ne k} {{{\mathbf{W}}_j}} } \right\} + {\mathrm{tr}}\left\{ {{{\mathbf{G}}_m}{{\mathbf{W}}_v}} \right\} + \sigma _m^2}}} \right)$.

\end{itemize}
	
To be clear, we summarize the difference for all the schemes in Table \ref{table4}.

\begin{table*}[t]
	\centering
	
	\caption{{Comparison of Schemes}}
	\label{table4}
	\begin{threeparttable}
		{
			\begin{tabular}{c|c|c|c|c|c}
				\Xhline{1.2pt}
				Scheme & Sensing signal & \makecell[c]{Related \\references} & $U_k$'s Rate  & $E_m$'s Rate & Power   \\
				\hline
				Scheme1 & {\makecell[c]{${s_v}$\\$\left( {{\alpha _1} = 0,{\alpha _2} = 1} \right)$}} & \cite{XiaH2023CL}  & \multirow{3}{*}{${ R_{c,k}}$ and ${R_{p,k}}$} & \multirow{3}{*}{${R_{c,m}^{\mathrm{E}}}$ \& ${R_{p,k,m}^{\mathrm{E}}}$ } & \multirow{3}{*}{${\mathrm{tr}}\left\{ {{{\mathbf{W}}_c} + {{\mathbf{W}}_v} + \sum\limits_{i = 1}^K {{{\mathbf{W}}_i}} } \right\}$}  \\
				\cline{1-3}
				Scheme2 & {\makecell[c]{${s_c}$\\$\left( {{\alpha _1} = 1,{\alpha _2} = 0} \right)$}} &   &  &  &   \\
				\cline{1-3}
				Scheme3 & {\makecell[c]{${s_c}\&{s_v}$\\$\left( {{\alpha _1} = 1,{\alpha _2} = 1} \right)$}} &   &  &  &  \\
				\hline
				Ben1 & ${s_c}$ & \cite{XuC2021JSTSP, YinY2022CL}  & \makecell[c]{${ R_{c,k}}$ and ${R_{p,k}}$ \\with ${{\mathbf{w}}_{v}} = 0$} & \makecell[c]{${R_{c,m}^{\mathrm{E}}}$ \& ${R_{p,k,m}^{\mathrm{E}}}$\\ with ${{\mathbf{w}}_{v}} = 0$} & ${\mathrm{tr}}\left\{ {{{\mathbf{W}}_c} + \sum\limits_{i = 1}^K {{{\mathbf{W}}_i}} } \right\}$  \\
				\hline
				SDMA & ${s_v}$ & \makecell[c]{\cite{ChenZ2024WCL}, \cite{LiuZ2024IOT}, \cite{ChenKX2024TVT},\\\cite{XiaH2024TWC}, \cite{ZhangC2024WCL}, \cite{LiuZ2024ARX}}  & ${R_{k,{\mathrm{SDMA}}}}$ & $R_{k,m,{\mathrm{SDMA}}}^{\mathrm{E}}$ & ${\mathrm{tr}}\left\{ {{{\mathbf{W}}_v} + \sum\limits_{i = 1}^K {{{\mathbf{W}}_i}} } \right\}$  \\
				\Xhline{1.2pt}
		\end{tabular}			}
	\end{threeparttable}
\end{table*}

\color{black}
	
\subsection{The Scenarios with Eavesdropping CSI}

\begin{figure}[t]
	\centering
	\subfigure[Algorithm 1.]{
		\label{fig02a}
		\includegraphics[width = 0.31 \textwidth]{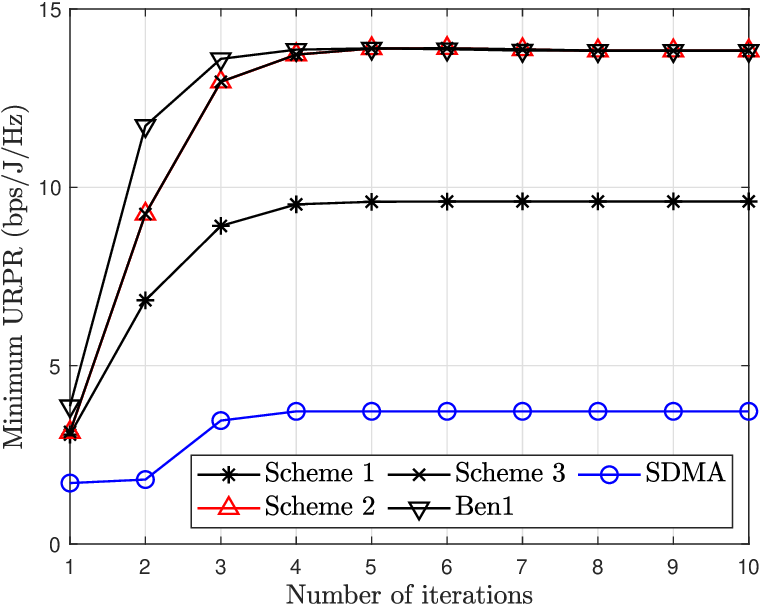}}
	\subfigure[Algorithm 2.]{
		\label{fig02b}
		\includegraphics[width = 0.31 \textwidth]{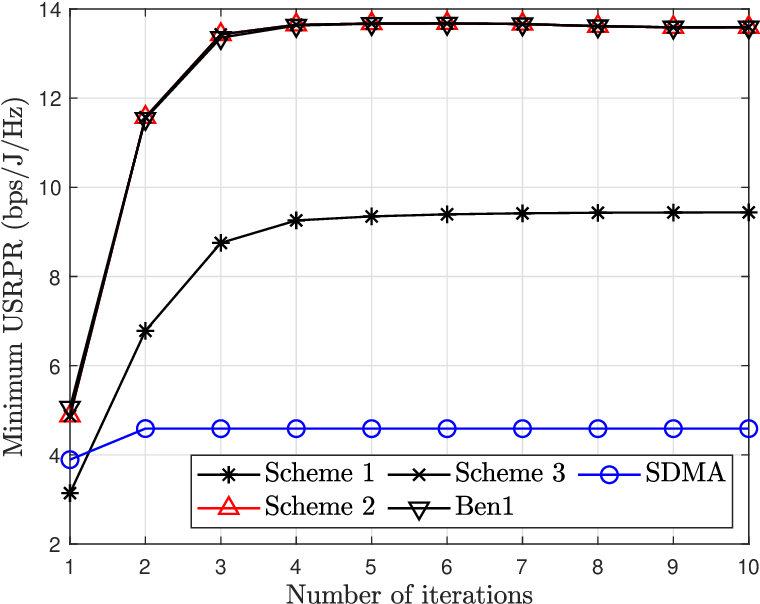}}	
	\caption{Convergence of Algorithms.}
	\label{fig02}
\end{figure}

Fig. \ref{fig02} presents the convergence of Algorithms 1 and 2. It can be observed that when the iteration reaches around 5, all algorithms tend to be stable, which indicates that the algorithms have a good convergence. 
Moreover, SDMA's performance is the worst, followed by Scheme 1's. Schemes 2 and 3 and Ben1 have better performance. 
The reason is that the common stream in Scheme 2 and Scheme 3 is reused for sensing. Compared with Scheme 1, thus, no additional power is required to fulfill the sensing requirements. Compared with SDMA, the SIC in RSMA (Schemes 2 and 3 and Ben1) can effectively improve the communication performance of the considered system.
Furthermore, in Figs. \ref{fig02a}-\ref{fig02b}, the performance of Schemes 2 and 3 and Ben1 is close. Since the difference among Scheme 2, Scheme 3, and Ben1 lies in which signal is used for sensing, the result demonstrates that \textbf{with the same power, there is no difference in sensing performance between using one signal and using two signals}. This conclusion is also demonstrated by all the following figures, and we omit Scheme 3 and the discussion due to the limited pages.

\begin{figure*}[t]
	\centering
	\subfigure[${{\mathbf{W}}_c}$.]{
		\label{fig03a}
		\includegraphics[width = 0.31 \textwidth]{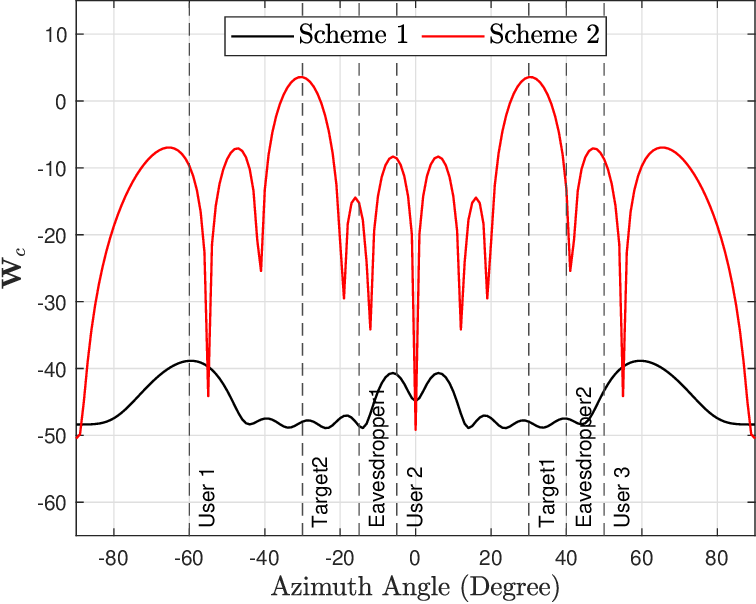}}
	\subfigure[${{\mathbf{W}}_1}$.]{
		\label{fig03b}
		\includegraphics[width = 0.31 \textwidth]{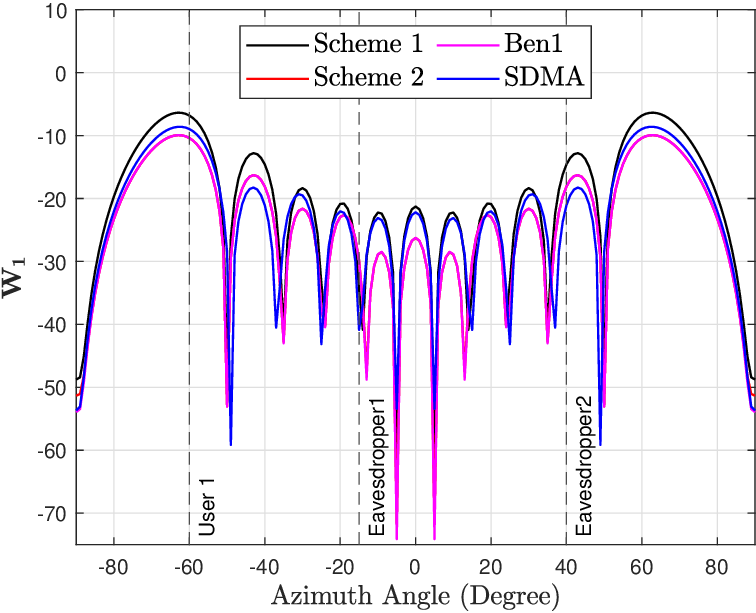}}	
	\subfigure[${{\mathbf{W}}_2}$.]{
		\label{fig03c}
		\includegraphics[width = 0.31 \textwidth]{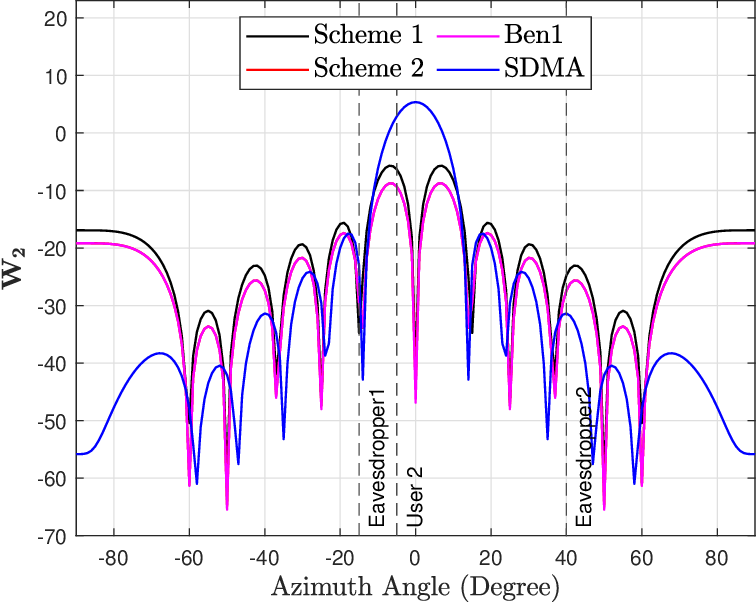}}
	\subfigure[${{\mathbf{W}}_3}$.]{
		\label{fig03d}
		\includegraphics[width = 0.31 \textwidth]{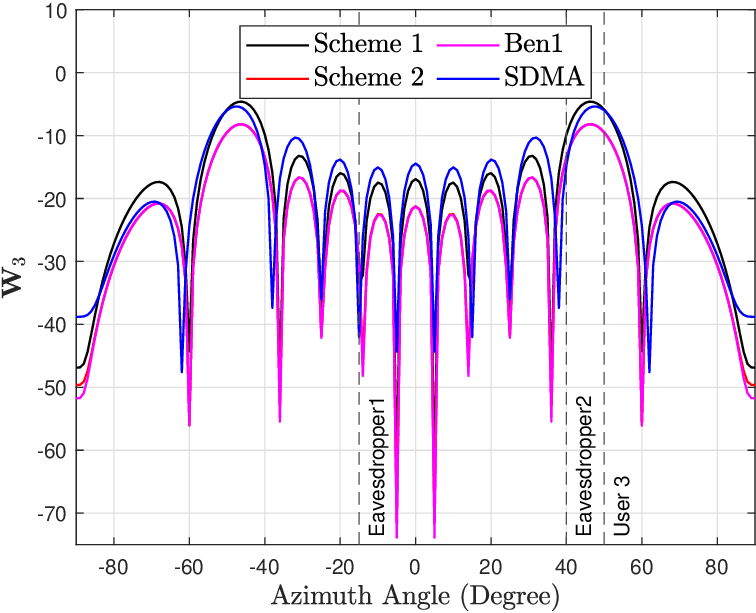}}
	\subfigure[${{\mathbf{W}}_v}$. ]{
		\label{fig03e}
		\includegraphics[width = 0.31 \textwidth]{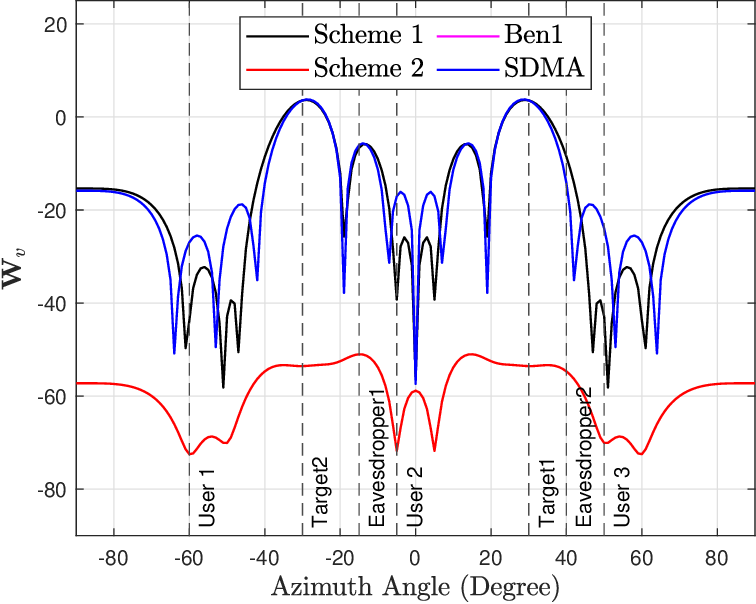}}
	\subfigure[Power allocation. ]{
		\label{fig03f}
		\includegraphics[width = 0.31 \textwidth]{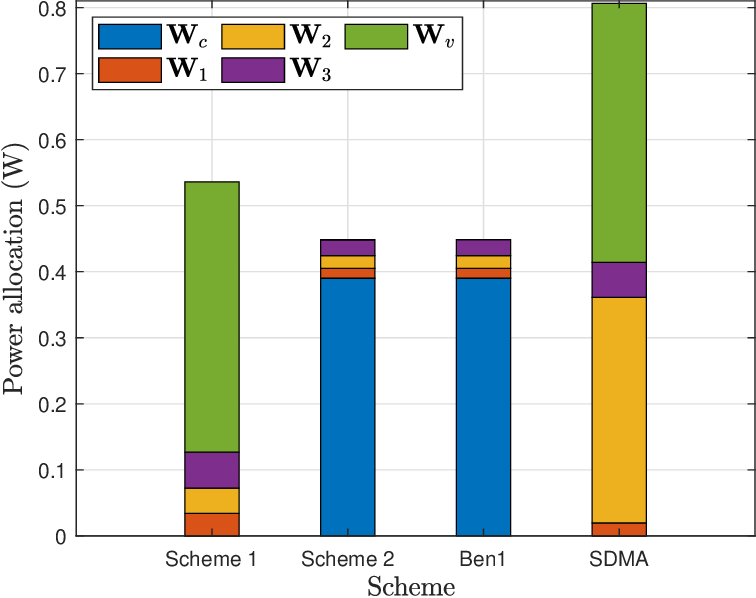}}
	\caption{Beamforming gains and power allocation in Algorithm 1 for different schemes.}
	\label{fig03}
\end{figure*}

Fig. \ref{fig03} plots the beamforming gain and power allocation with all the schemes.
Fig. \ref{fig03a} presents the beamforming gains of the common stream. As can be seen, Schemes 1 and 2 have main lobes in the users' direction and targets' direction, respectively. This is because the common stream in Scheme 1 is utilized for communication, while the common stream in Scheme 2 is utilized for both communication and sensing simultaneously.
Figs. \ref{fig03b}-\ref{fig03d} respectively plot the beamforming gain of ${{\mathbf{w}}_1}$, ${{\mathbf{w}}_2}$, and ${{\mathbf{w}}_3}$. It is clearly seen from the figures that the two proposed schemes have slightly higher and lower gains, respectively, in the LUs' direction and the eavesdroppers' direction to ensure the security constraint (\ref{P1e}) of the private data stream. 
Fig. \ref{fig03e} presents the beamforming gain of ${{\mathbf{w}}_v}$. We can find that both SDMA and Scheme 1 have the main lobes in the targets' direction because ${{\mathbf{s}}_v}$ is utilized for sensing in both schemes. However, in Scheme 2, ${{\mathbf{s}}_v}$ is regarded as AN, so the gain in the LUs' direction is relatively lower.
Fig. \ref{fig03f} shows the power allocation with different schemes. It can be clearly seen from the figure that SDMA consumes the most significant power, while Scheme 2 and Ben1 consume the least power. Furthermore, the proportion of sensing in all the schemes is the largest. In Scheme 2, the common stream is multiplexed to sense during the process of information transmission, thereby enhancing the energy efficiency of the considered RSMA system. 
Comparison of Figs. \ref{fig03a} and \ref{fig03e}, it can be observed that the BF gain for the AN (${{\mathbf{W}}_c}$ in Scheme 1 and ${{\mathbf{W}}_v}$ in Scheme 2) is lower, which testifies that compared with BF, AN is limited in improving system performance. The same conclusion also can be found in Fig. \ref{fig03f} by comparing the power allocated for 
${{\mathbf{w}}_c}$ in Scheme 1 and ${{\mathbf{w}}_v}$ in Scheme 2.
	
\begin{figure*}[t]
	\centering
	\subfigure[Algorithm 1.]{
		\label{fig04a}
		\includegraphics[width = 0.22 \textwidth]{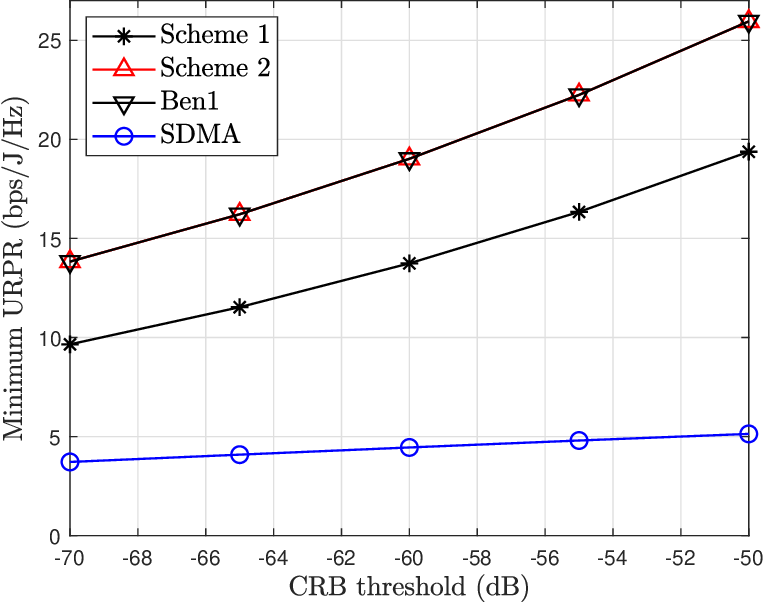}}
	\subfigure[Algorithm 2.]{
		\label{fig04b}
		\includegraphics[width = 0.22 \textwidth]{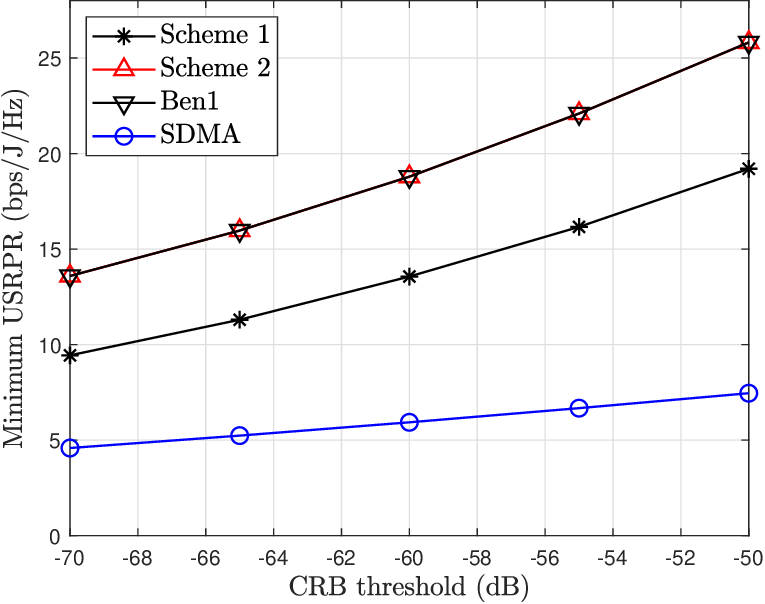}}	
	\subfigure[Algorithm 1 with $\vartheta= -60$ dB.]{
		\label{fig04c}
		\includegraphics[width = 0.22 \textwidth]{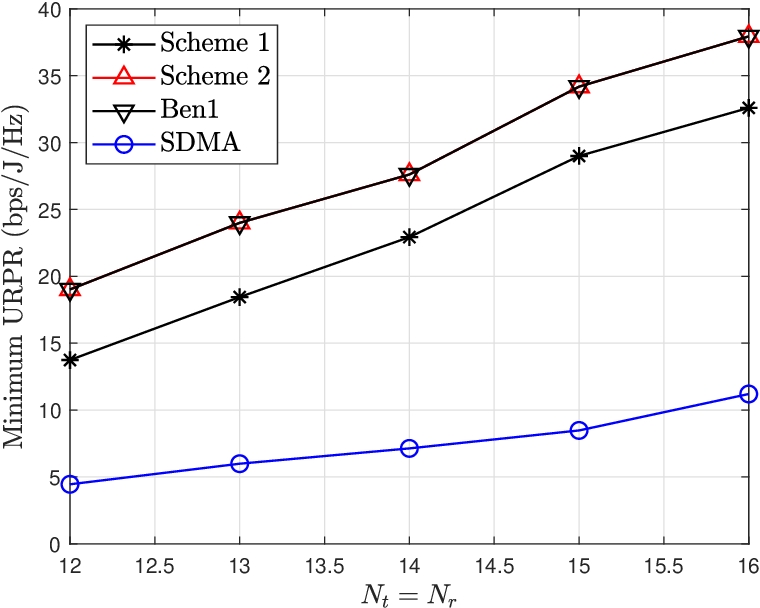}}
	\subfigure[Algorithm 2 with $\vartheta= -60$ dB.]{
		\label{fig04d}
		\includegraphics[width = 0.22 \textwidth]{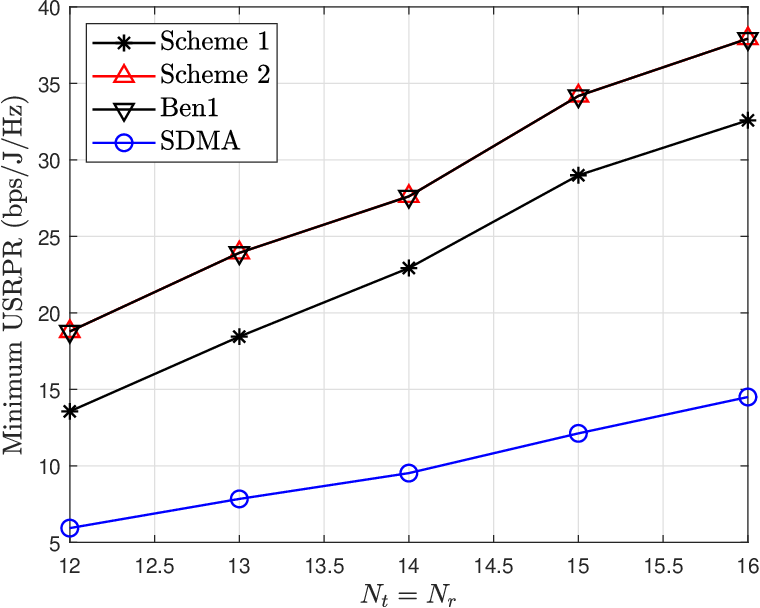}}	
	\caption{Performance versus varying CRB threshold or $N_t = N_r$.}
	\label{fig04} 
\end{figure*}
	
Figs. \ref{fig04a}-\ref{fig04d} depict the communication performance versus varying CRB threshold and $N_t = N_r$, respectively. 
It can be seen from Figs. \ref{fig04a} and \ref{fig04b} that the minimum URPR and USRPR increase and the minimum NPC decreases with the increase of the given threshold. This is because the larger the CRB threshold, the weaker the demand for sensing; thereby, more power would be allocated to communication. 
The extra signal is utilized in SDMA to sense, which can also interfere with LUs. When the sensing requirement decreases, the interference from extra signals reduces, thus increasing communication performance. 
Moreover, the performance of Scheme 2 is better than that of Scheme 1. This is because the common stream (${{\mathbf{s}}_c}$) and extra signal (${{\mathbf{s}}_v}$) are utilized in Scheme 2 and Scheme 1, respectively. 
It should be noted that ${{\mathbf{s}}_c}$ can be decoded and ${{\mathbf{s}}_v}$ would not be decoded in all the LUs, given in (\ref{barSNRpk}), 
which denotes SIC in RSMA systems can enhance the communication performance.
Figs. \ref{fig04c} and \ref{fig04d} show that as the number of antennas increases, the communication performance is improved. This is because the increase in the number of antennas enhances the channel capacity and degrees of freedom. The performance growth of Schemes 1 and 2 outperforms that of SDMA, which demonstrates that RSMA has an advantage over SDMA in multi-antenna scenarios.

\begin{figure*}[t]
	\centering
	\subfigure[Algorithm 1.]{
		\label{fig05a}
		\includegraphics[width = 0.31 \textwidth]{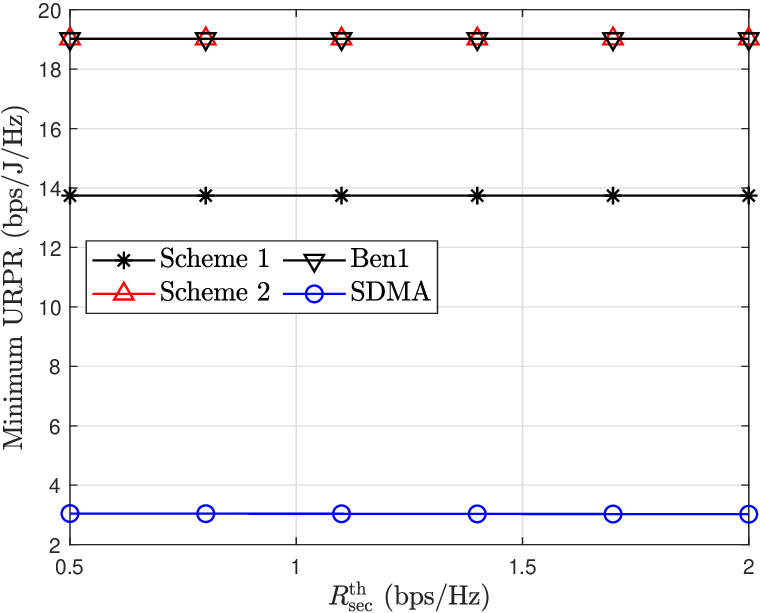}}
	\subfigure[Algorithm 2.]{
		\label{fig05b}
		\includegraphics[width = 0.31 \textwidth]{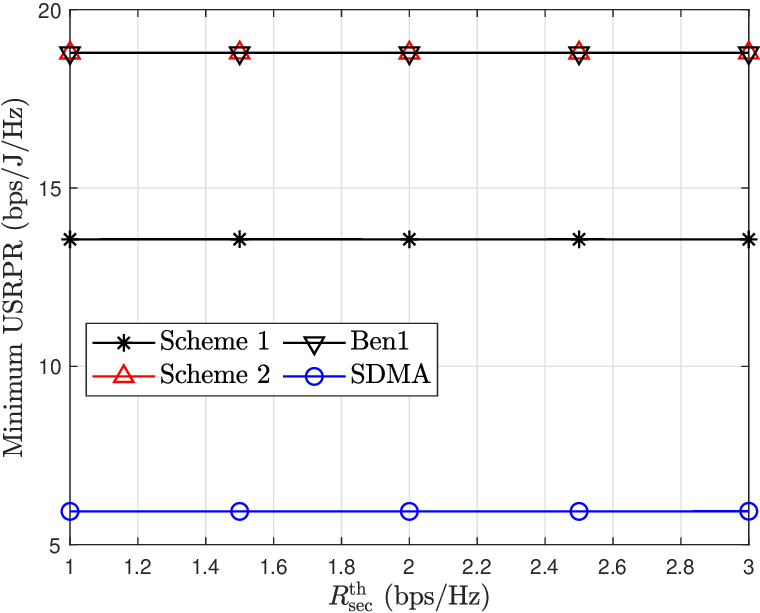}}	
	\subfigure[Algorithm 1.]{
		\label{fig05c}
		\includegraphics[width = 0.31 \textwidth]{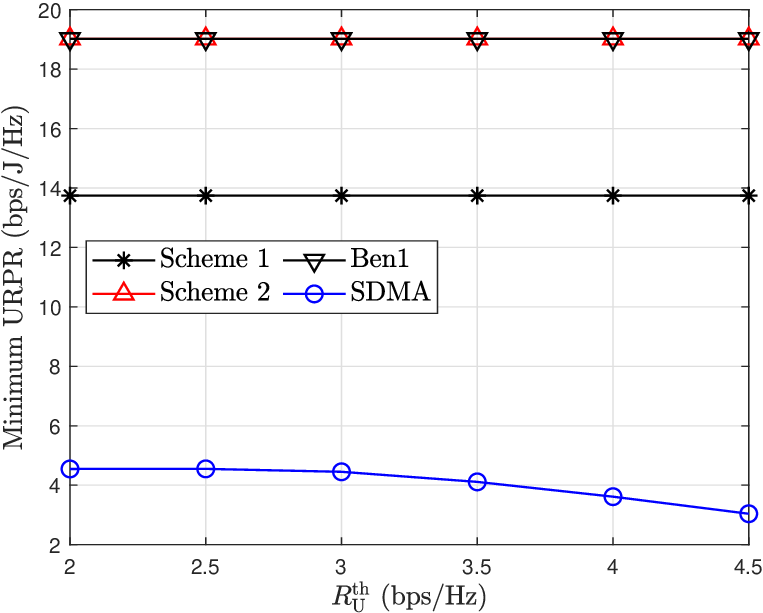}}
	\caption{Performance versus varying $R_{\sec }^{\mathrm {th}}$ or $R_{{\mathrm{U}}} ^{{\mathrm {th}}}$ with $\vartheta= -60$ dB.}
	\label{fig05}
\end{figure*}

Figs. \ref{fig05a} and \ref{fig05b} depict the minimum UPRP and minimum USPRP corresponding to varying secrecy rate thresholds. 
It can be seen that the effect of the security rate threshold on the URPR and USRPR can be ignored. 
This is because increasing the secrecy rate threshold denotes higher security requirements, which leads to the requirement for more power.
Fig. \ref{fig05c} represents the minimum UPRP corresponding to the QoS threshold. 
With the increase of the QoS threshold, the URPR of Schemes 1 and 2 remains constant, while that of SDMA decreases in the high-requirement region. This is because SDMA needs more power for the same requirement.	

\color{black}
\subsection{The Scenarios without Eavesdropping CSI}

\begin{figure*}[t]
	\centering
	\subfigure[]{
		\label{fig06a}
		\includegraphics[width = 0.22 \textwidth]{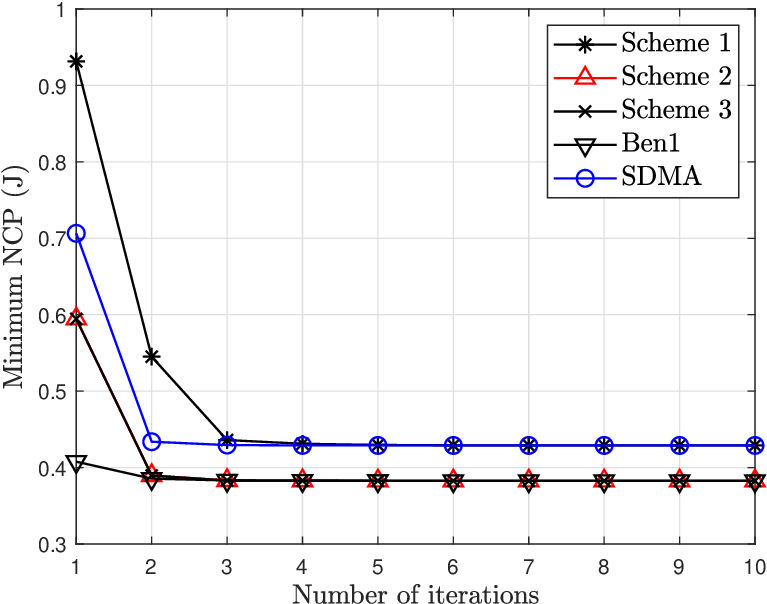}}
	\subfigure[]{
		\label{fig06b}
		\includegraphics[width = 0.22 \textwidth]{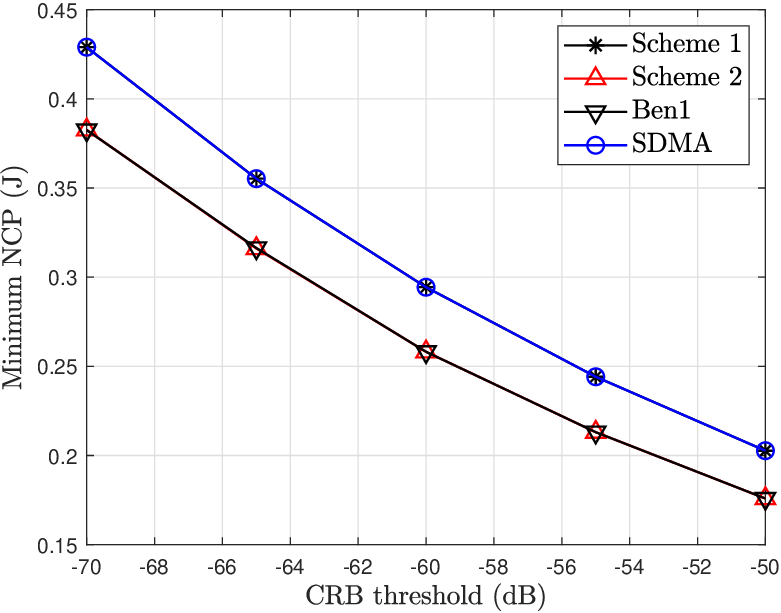}}
	\subfigure[]{
		\label{fig06c}
		\includegraphics[width = 0.22 \textwidth]{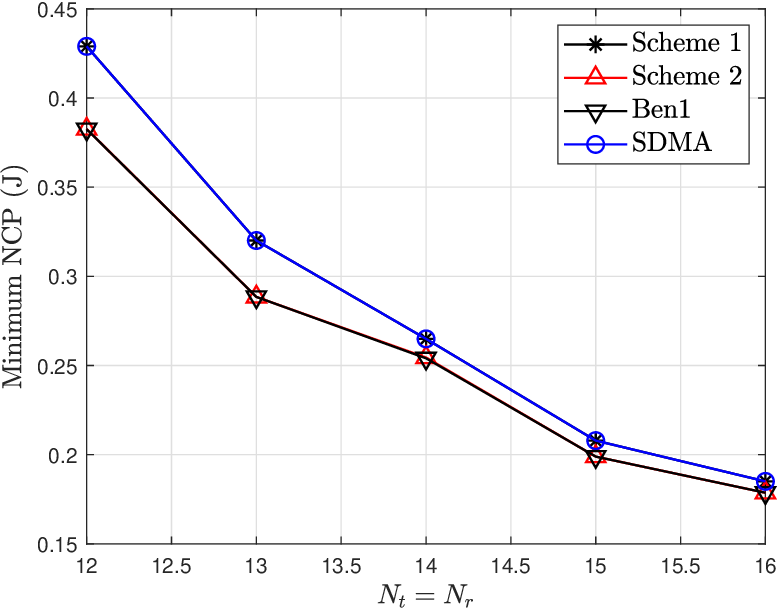}}	
	\subfigure[]{
		\label{fig06d}
		\includegraphics[width = 0.22 \textwidth]{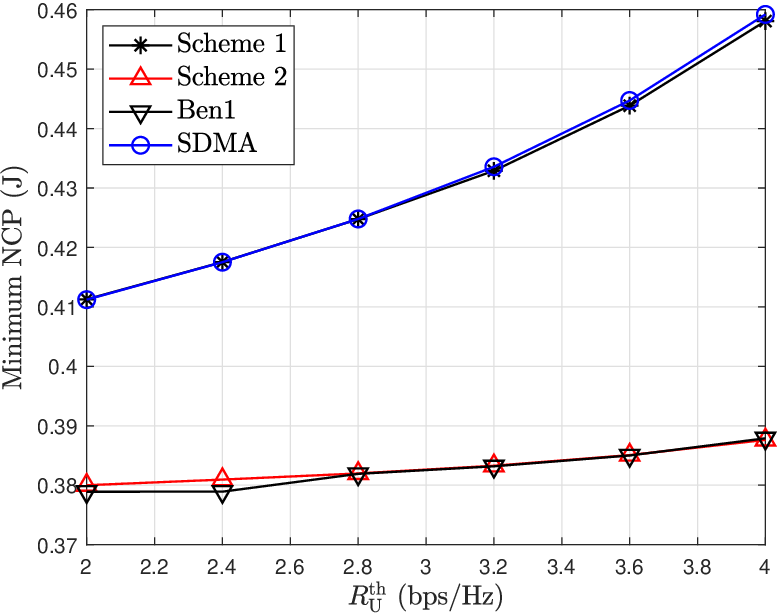}}
	\caption{(a) Convergence of Algorithm3. Performance of Algorithm 3 versus varying (b) CRB threshold. (c) $N_t = N_r$. (d) QoS requirement with $P_{\max} = 40$ dBm. }
	\label{fig06}
\end{figure*}

Figs. \ref{fig06a}-\ref{fig06d} present the convergence of Algorithms 3, the minimum NPC versus varying CRB threshold, $N_t = N_r$, QoS requirement. It can be observed from Fig. \ref{fig06a} that Algorithm 3 also has good convergence, like Algorithms 1 and 2. 
In addition to some conclusions similar to those in Section V.A  (for example, Schemes 2 and 3 and Ben1 have the same performance, \textit{etc.}), one interesting conclusion can be found that the performance of Scheme 1 is comparable to that of SDMA. 
This is because for scenarios without eavesdropping CSI, like SDMA, Scheme 1 also uses extra signals to complete the sensing and thus cannot reduce the NPC of the considered system.
Furthermore, the difference between SDMA and Scheme 1 is the common stream in Scheme 1 is utilized as AN, and the results demonstrate that \textbf{compared with BF, the effect of AN on improvement security can be ignored}, which also is demonstrated in Figs. \ref{fig06b}-\ref{fig06d}.

\color{black}
													
\section{Conclusion}
\label{sec:Conclusions}
														
This work investigated the performance of RSMA-assisted ISAC systems in multi-user scenarios involving multiple LUs, eavesdroppers, and sensing targets. Two distinct approaches were examined: one utilizing dedicated sensing signals and another employing the common stream for sensing. Both scenarios with and without eavesdropping CSI were considered. To address the system design challenges, we jointly optimized BFs and CRA to minimize the transmit power at the ISAC BS while simultaneously maximizing the minimum achievable rate among all LUs. To solve the resulting non-convex optimization problem, we developed efficient iterative algorithms based on SCA techniques, which systematically transform non-convex constraints into convex forms. 
{
	In this work, it is assumed that the perfect SIC and CSI of LUs is known at BS. Utilizing the bounded error model \cite{FuH2020TWC, ZhangC2024WCL} and random error model \cite{LoliRC2022ARX, DizdarO2022OJCS, XiaH2024TWC}, considering the imperfect CSI and SIC will be an interesting work and will be part of future work.
	Moreover, to the best of our knowledge, there are mmWave-enabled TD-ISAC hardware testbed \cite{ZhangQ2024Net} and RSMA prototype \cite{LyuX2024TWC}. Based on these hardware platforms, developing a new test platform and testing the performance of the proposed schemes will be part of future work.
}

\begin{appendices}
	
\section{}
\label{sec:appendicesA}	

For a concave function of $f\left( {\mathbf{X}} \right)$ with respect to ${\mathbf{X}}$, its first-order Taylor's formula is expressed as
\begin{align}
	f\left( {\mathbf{X}} \right) \le f\left( {{\mathbf{\tilde X}}} \right) + {\mathrm{vec}}\left( {f'\left( {{\mathbf{\tilde X}}} \right)} \right){\mathrm{vec}}\left( {{\mathbf{X}} - {\mathbf{\tilde X}}} \right),
\end{align}
where 
${{\mathbf{\tilde X}}}$ denotes a given feasible point. 
Based on $\partial \left( {\log \left( {{\mathrm{tr}}\left( {\mathbf{X}} \right)} \right)} \right) = {\left( {{\mathrm{tr}}\left( {\mathbf{X}} \right)} \right)^{ - 1}}{\mathrm{tr}}\left( {\partial {\mathbf{X}}} \right)$, 
the upper bound of log function is obtained as 
(\ref{eqlast}), shown at the top of this page.
\begin{figure*}[ht]
	\hrulefill
	\begin{align}
		{\log _2}\left( {{\mathrm{tr}}\left( {{\sigma ^2}{\mathbf{I}} + {\mathbf{HW}}} \right)} \right) &\le {\log _2}\left( {{\mathrm{tr}}\left( {{\sigma ^2}{\mathbf{I}} + {\mathbf{H\hat W}}} \right)} \right) + \frac{1}{{\ln 2}}{\mathrm{vec}}\left( {{{\left( {{\sigma ^2}{\mathbf{I}} + {\mathbf{H\hat W}}} \right)}^{ - 1}}{\mathbf{H}}} \right){\mathrm{vec}}\left( {{\mathbf{W}} - {\mathbf{\hat W}}} \right) \nonumber\\
		&= {\log _2}\left( {{\mathrm{tr}}\left( {{\sigma ^2}{\mathbf{I}} + {\mathbf{H\hat W}}} \right)} \right) + \frac{1}{{\ln 2}}\left( {{{\left( {{\mathrm{tr}}\left( {{\sigma ^2}{\mathbf{I}} + {\mathbf{H\hat W}}} \right)} \right)}^{ - 1}}{\mathrm{tr}}\left( {{\mathbf{HW}}} \right)} \right) \nonumber\\
		&- \frac{1}{{\ln 2}}\left( {{{\left( {{\mathrm{tr}}\left( {{\sigma ^2}{\mathbf{I}} + {\mathbf{H\hat W}}} \right)} \right)}^{ - 1}}{\mathrm{tr}}\left( {{\mathbf{H\hat W}}} \right)} \right)
		\label{eqlast}
	\end{align}
	\hrulefill
\end{figure*}

\end{appendices}

\end{document}